\documentclass[final,3p,times,onecolumn, sort&compress]{elsarticle}

\usepackage{lineno,hyperref}
\modulolinenumbers[5]

\usepackage{amsmath}
\usepackage{amsfonts,color}
\usepackage{amssymb,float}
\usepackage{mathtools}
\usepackage[utf8]{inputenc}
\usepackage{graphicx}
\usepackage{wasysym}
\usepackage{tabularx}
\usepackage{accents}
\usepackage{graphicx}
\usepackage{color, colortbl}
\usepackage[dvipsnames]{xcolor}
\usepackage{hyperref}
\usepackage{ulem}

\hypersetup{
    colorlinks=true,
    linkcolor=blue,
    filecolor=magenta,      
    citecolor=red
}

\usepackage{subfigure} 
\usepackage{hyperref} 
\usepackage{multirow}

\definecolor{Gray}{gray}{0.9}
\definecolor{LightCyan}{rgb}{0.88,1,1}

\newcommand{\be}{\begin{equation}}
\newcommand{\ee}{\end{equation}}
\newcommand{\bea}{\begin{eqnarray}}
\newcommand{\eea}{\end{eqnarray}}

\journal{}

\begin{document}

\begin{frontmatter}

\title{Hubble constant by natural selection: Evolution chips in the Hubble tension}


\author[a]{Reginald Christian Bernardo\corref{mycorrespondingauthor}}
\cortext[mycorrespondingauthor]{Corresponding author}
\ead{rbernardo@gate.sinica.edu.tw}

\author[b]{You-Ru Lee}
\ead{00851043@email.ntou.edu.tw}

\address[a]{Institute of Physics, Academia Sinica, Taipei 11529, Taiwan}
\address[b]{Department of System Engineering, National Taiwan Ocean University, Keelung 202301, Taiwan}

\begin{abstract}
The Approximate Bayesian Computation (ABC) algorithm considers natural selection in biology as a guiding principle for statistical model selection and parameter estimation. We take this ABC approach to cosmology and use it to infer which $\Lambda$CDM model anchored on a choice of a Hubble constant prior would be preferred by the data. We find in all of our runs that the Planck Hubble constant ($H_0 = 67.4 \pm 0.5$ km s$^{-1}$Mpc$^{-1}$) always emerge naturally selected by the ABC over the SH$0$ES estimate ($H_0 = 73.30 \pm 1.04$ km s$^{-1}$Mpc$^{-1}$). The result holds regardless of how we mix our data sets, including supernovae, cosmic chronometers, baryon acoustic oscillations, and growth data. Compared with the traditional MCMC, we find that the ABC always results with narrower cosmological constraints, but remain consistent inside the corresponding MCMC posteriors.
\end{abstract}


\end{frontmatter}




\section{Introduction}
\label{sec:intro}

The standard cosmological model, $\Lambda$CDM, is parametrically the best model of the Universe \cite{Peebles:2002gy,Copeland:2006wr}, providing a reliable picture of the evolution \cite{Riess:1998cb,Perlmutter:1998np} and the growth and dynamics of matter in it \cite{Guth:1980zm,Linde:1981mu,Bertone:2004pz,Baudis:2016qwx}. However, it glosses over several theoretical issues such as the fine tuning of the cosmological constant \cite{RevModPhys.61.1} and the coincidence \cite{Dalal:2001dt} problem, and recently it was recognized that it caters a tension in its cosmological parameters \cite{DiValentino:2020vhf,DiValentino:2020zio,DiValentino:2020vvd}. The Hubble tension \cite{Schoneberg:2021qvd} -- a three to five sigma deviation between the local distance ladder measurements \cite{Riess:2021jrx} and cosmic microwave background \cite{Aghanim:2018eyx} estimates of the Hubble constant -- raises serious considerations to noncanonical statistical methods \cite{Escamilla-Rivera:2020fxq, Briffa:2020qli, LeviSaid:2021yat, Mukherjee:2022yyq, Benisty:2022psx, Grandon:2022gdr}, dynamical dark energy \cite{Cardenas:2014jya, Vagnozzi:2019ezj, Grandon:2021nls, Bernardo:2021cxi}, and even cosmological models with additional degrees of freedom in the matter and gravitational sectors \cite{Capozziello:2011et, Peirone:2019aua, Frusciante:2018jzw,  Bahamonde:2019shr, Bernardo:2021qhu, Odintsov:2022eqm}. We take the former route in this paper.

The Approximate Bayesian Computation (ABC) is a parameter estimation algorithm that brings forth the concept of natural evolution to the physical parameters in a model \cite{10.1093/genetics/162.4.2025, 2009arXiv0901.1925T}. The main novelty of this method is that it bypasses the concept of a likelihood, and rather instead depends on a distance function. This is a useful feature particularly in model selection where the likelihood between different models is often unknown or at best intractable. Together with a sequential Monte Carlo spin that lets only the finest of a population proceed to the next generation, the ABC becomes even more powerful, resulting to only an ever sharpening constraints on the models and parameters during the course of the evolution \cite{2009arXiv0901.1925T}. The algorithm has been utilized for model selection and parameter estimation in various scientific problems \cite{10.2307/27896228, 10.3389/fbuil.2017.00052,  10.1093/bioinformatics/bty361} and in astrophysics and cosmology \cite{2013ApJ...764..116W, Akeret:2015uha, Ishida:2015wla, 2017A&C....19...16J, Bernardo:2021mfs, Bernardo:2022pyz, Kilbinger:2021dxv}.

In this work, we make use of the ABC to answer an interesting question in the background of the Hubble tension: ``Which Hubble constant is preferred by the present data, SH$0$ES \cite{Riess:2021jrx} or Planck \cite{Aghanim:2018eyx}?'' In this endeavor, we consider the standard cosmological model, $\Lambda$CDM, together with cosmological observations of the late Universe. We anchor $\Lambda$CDM on two Hubble constant priors that reflect the Hubble tension, and let the ABC tell which of the two resulting models is preferred by natural selection. This provides independent estimates of the cosmological parameters of the standard model, but more importantly openly views the Hubble tension in a new light. With such newfound statistical perspective, we find the results of the ABC even more appealing as one Hubble constant emerges throughout, regardless of the priors and the observations in use. We also briefly discuss the results of the traditional Markov chain Monte Carlo (MCMC) analysis \cite{Trotta:2008qt, Lewis:2019xzd, 2020arXiv200505290T} for the same data sets, setting a baseline for the comparison of ABC as a statistics tool for cosmology.

Our goal in this work is modest: introduce the ABC as a model selection tool for cosmology, through its take on the Hubble tension, arguably cosmology's biggest controversy today. We acknowledge that the Hubble constants we consider in this work have differing nature, in the sense that the local measurement \cite{Riess:2021jrx} give the present expansion rate directly, while that from early Universe \cite{Aghanim:2018eyx} rely on a physical model to extrapolate by a billion years to meet with the present cosmological conditions. In retrospect, comparing these through ABC as a prior is admittedly superficial, bypassing a lot of obviously exciting cosmological details, but it achieves our demonstrative purpose which we hope gives ABC more attention as an equally powerful tool for cosmological analysis, surrogate to the canonical MCMC when often a likelihood is unavailable.

The rest of this paper is as follows. We provide an overview of the ABC (Section \ref{sec:abc}) and then a brief description of the data sets and how they were used in the analysis (Section \ref{sec:data_sets}). We then present our main results using MCMC and ABC (Section \ref{sec:results}){, discuss the robustness of the ABC in terms of the prior dependence (Section \ref{sec:varyingH0priors}),} and draw our conclusions (Section \ref{sec:conclusions}). {\ref{sec:matterABC} and \ref{sec:abcsmc} give details to ABC constraints on the matter density, and provide explicit steps of the algorithm.}

{We share our python notebooks in \href{https://github.com/reggiebernardo/notebooks/tree/main/supp_ntbks_arxiv.2212.02203}{GitHub} so that the readers may at their own leisure further indulge themselves in the details and the enjoyment that come with using ABC/MCMC. We acknowledge the use of python packages, pyABC \cite{pyabc} for ABC implementation, Cobaya \cite{2020arXiv200505290T} for MCMC, and GetDist \cite{Lewis:2019xzd} for statistical analysis.}

\section{The ABC method}
\label{sec:abc}

Approximate Bayesian computation, in short ABC \cite{10.1093/genetics/162.4.2025}, is applicable to finding the posterior probability $P\left(\theta|D\right)$ for the parameters $\theta$ given the data $D$, by combining the prior $P(\theta)$ and observed data to train the model so that the constrained model parameters can be obtained based on Bayes' theorem. This overcomes the traditional MCMC based Bayesian inference, which relies pragmatically on the likelihood function $P(D|\theta)$ between the observation and data, which is often challenging or intractable especially for complex stochastic systems such as in population genetics. The ABC was produced to provide an alternative likelihood-free route of statistically analyzing systems, thereby simplifying the process and improving its efficiency, and making it applicable to a wider range of scientific problems such as in biology \cite{2009arXiv0901.1925T, 10.2307/27896228, 10.3389/fbuil.2017.00052, 10.1093/bioinformatics/bty361} and physics \cite{2013ApJ...764..116W, Akeret:2015uha, Ishida:2015wla, 2017A&C....19...16J, Bernardo:2021mfs, Bernardo:2022pyz}.

The main idea behind ABC is to approximate the posterior, $P(\theta|D) \propto f\left(D^*, D\right) P(\theta)$, by building on distance function $f\left(X, Y\right)$ measurements of a population, where each member is drawn based on the model, providing its own simulated data $D^*$. Through large population sizes and an acceptance criterion, $f\left(D^*, D\right) < \epsilon$ for some sufficiently small $\epsilon$, the ABC is able to obtain a competitive estimate of the model parameters that rival those from MCMC. The algorithm receives an even more significant power up when given a sequential Monte Carlo (SMC) twist \cite{2009arXiv0901.1925T}. This makes an overall process that estimates the parameters in an inevitably tightening sequence of generations. Since the ABC is likelihood free, all these can be put together in the larger space that covers even models as statistical points, that is a bird eye view that sees models with their own probability, while also measuring their parameters. As a consequence, the resulting ABC-SMC algorithm is a natural model selection and parameter estimation statistical tool. Needless to say, it is an attractive one, free from the restrictions that a likelihood incorporates into the analysis. For this reason it is understandably applied to a variety of problems transcending scientific fields \cite{10.1093/genetics/162.4.2025, 10.2307/27896228, 2013ApJ...764..116W, Akeret:2015uha, Ishida:2015wla, 2017A&C....19...16J, 10.1093/bioinformatics/bty361, Bernardo:2021mfs, Bernardo:2022pyz}.

{
We refer the reader to \cite{2009arXiv0901.1925T, 2009arXiv0911.1705T, 2009arXiv0910.4472T} for an outstanding introduction to ABC (\ref{sec:abcsmc}), and to the python package pyABC \cite{10.1093/bioinformatics/bty361} for an implementation of the algorithm complemented with many examples.
}

Building on several work that has now used ABC in astrophysics and cosmology, we want to see ABC's take on the Hubble tension in the standard cosmological model by making use of two values that represent it as a parameter prior. To accomplish this, we utilize a python package `pyABC' \cite{pyabc}, which is known for its ABC-SMC feature, allowing model selection and parameter estimation by simply inputting the models, observed data, the parameter priors, and a distance function, taking the place of the likelihood in order to sequentially approximate the posterior throughout the evolution. In the end, we have only one model we deem naturally selected.

\section{Cosmological data sets}
\label{sec:data_sets}

We consider background cosmological data from cosmic chronometers \cite{2010JCAP...02..008S, 2012JCAP...08..006M, 2014RAA....14.1221Z, Moresco:2015cya, Moresco:2016mzx, Ratsimbazafy:2017vga}, supernovae \cite{Brout:2021mpj, Brout:2022vxf, Scolnic:2021amr}, and baryon acoustic oscillations \cite{Gaz34, 2012MNRAS.425..405B, Chuang:2013hya, Busca36, BOSS:2013igd, Anderson32, Oka37, BOSS:2014hwf, Wang33, Bautista:2017zgn, Alam38}. The expansion rate data is compiled in the Appendix of \cite{Bernardo:2021cxi} for convenience. We also consider two independent Hubble constant measurements that represent the Hubble tension \cite{Aghanim:2018eyx, Riess:2021jrx}. In addition, we take in the growth data from redshift space distortions for cosmological perturbations compiled in \cite{Kazantzidis:2018rnb}. We provide a brief description of each of these data and how we used it for our analysis below.

\begin{itemize}
\item Cosmic chronometers (CC). The cosmic chronometers which act as standard clocks of our universe provide a cosmology model independent way of measuring the expansion rate and the Hubble constant. This is based on observations of the age and metalicity of temporally adjacent passive galaxies such that the expansion rate at a redshift $z_{\rm CC}$ can be approximated by $H_{\rm CC}(z) \approx -\left(\Delta z_{\rm CC}/\Delta t\right)/\left(1 + z_{\rm CC}\right)$. CC observations at different redshifts therefore give a direct measurement of the expansion history which we use to constrain cosmology. We consider the CC in the redshift range $0.07 \lesssim z \lesssim 1.97$ from various sources \cite{2010JCAP...02..008S, 2012JCAP...08..006M, 2014RAA....14.1221Z, Moresco:2015cya, Moresco:2016mzx, Ratsimbazafy:2017vga} that is compiled in \cite{Bernardo:2021cxi} for a quick reference. In this work, we consider {a flat} $\Lambda$CDM cosmological model in which the expansion rate is provided as
\begin{equation}
    E\left(z \right)^2 = \Omega_{m0} \left(1 + z \right)^3 + \left(1 - \Omega_{m0}\right)
\end{equation}
where $E(z) = H(z)/H_0$ is the normalized Hubble expansion rate at redshift $z$, $H_0$ is the Hubble constant, and $\Omega_{m0}$ is the fraction of nonrelativistic matter at redshift $z = 0$. We compare the predicted expansion rate with the CC at different redshifts by the chi-squared distance
\begin{equation}
\chi^2_{\rm CC} = \sum_{z_{\rm CC}} \left( \dfrac{H\left(z_{\rm CC}\right) - H_{\rm CC}\left(z_{\rm CC}\right)}{\sigma_{\rm CC}\left(z_{\rm CC}\right)} \right)^2 \,,
\end{equation}
where the quantities with the subscript CC refer to the CC measurements of the expansion rate, for example, at redshift $z_{\rm CC}$ the expansion rate is $H_{\rm CC}\left(z_{\rm CC}\right) \pm \sigma_{\rm CC}\left(z_{\rm CC}\right)$. We consider uncorrelated CC observations.

\item Supernovae (SNe). We consider supernovae type Ia, so called standard candles of the Universe as they are intrinsically very bright that they serve as cosmic lampposts that can be used to estimate the distance from the observer. Flux measurements of SNe supplemented with redshift observations thus provide a measure of the expansion rate. We understand this in terms of the distance redshift ladder relation,
\begin{equation}
    m(z) - M = 5 \log_{10}\left( d_L\left(z\right) \right) \,,
\end{equation}
where $d_L(z)$ is the luminosity distance given by
\begin{equation}
    d_L(z) = \dfrac{c}{H_0} \left(1 + z\right) \int_0^z \dfrac{dz'}{E(z')} \,.
\end{equation}
We use the data set released by the Pantheon$+$ team \cite{Brout:2021mpj, Brout:2022vxf, Scolnic:2021amr} where there are $1701$ data points from type Ia supernovae. This gives apparent magnitude observations at redshift ranges $0.01 \lesssim z \lesssim 2.3$\footnote{We opt to use this conservative redshift range where the Pantheon+ samples are free from other tensions raised \cite{Pasten:2023rpc, Perivolaropoulos:2023iqj}.} together with a covariance matrix to take into account systematic influences. We compare the model predictions with SNe observations by treating the chi-squared measure as a distance function
\begin{equation}
    \chi^2_{\rm SNe} = \sum_{z_{\rm SNe}} \left( m(z_{\rm SNe}) - m_{\rm SNe}(z_{\rm SNe}) \right) C^{-1}_{\rm SNe} \left( m(z_{\rm SNe}) - m_{\rm SNe}(z_{\rm SNe}) \right) \,,
\end{equation}
where subscripts SNe stand for the Pantheon$+$ supernovae apparent magnitude measurements $\mu_{\rm SNe}$ with the covariance matrix $C_{\rm SNe}$.

\item Baryon acoustic oscillations (BAO). 
The baryon acoustic oscillations, which represent acoustic fluctuations in the early Universe, serve as standard rulers in the sky that is the distance the wave travels in the plasma before recombination. However, at a redshift $z$, this only directly measures the parameter combination $r_{\rm D}H(z)$ where the sound horizon radius $r_{\rm D}$ of the drag epoch is a cosmology independent quantity. Nonetheless since we are only concerned with $\Lambda$CDM this is given by $r_{\rm D} = 147.74$ Mpc, which then provides indirectly BAO measurements of the expansion rate at various redshifts \cite{Gaz34, 2012MNRAS.425..405B, Chuang:2013hya, Busca36, Chuang28, Oka37, Anderson32, BOSS:2013igd, BOSS:2014hwf, Bautista:2017zgn, Wang33, Alam38}. This supplements $26$ more data points to the cosmological expansion rate in the redshifts between $0.24 < z < 2.36$ (summarized in \cite{Bernardo:2021cxi}). As with CC, we measure the deviation of our predictions from observation by means of the distance function
\begin{equation}
\chi^2_{\rm BAO} = \sum_{z_{\rm BAO}} \left( \dfrac{H\left(z_{\rm BAO}\right) - H_{\rm BAO}\left(z_{\rm BAO}\right)}{\sigma_{\rm BAO}\left(z_{\rm BAO}\right)} \right)^2 \,,
\end{equation}
where the BAO measurements are given as $H_{\rm BAO}\left(z_{\rm BAO}\right) \pm \sigma_{\rm BAO}\left(z_{\rm BAO}\right)$, after dividing by the drag sound horizon radius.

\item Hubble constant measurements. In the background of the Hubble tension, we make use of the Hubble constant estimates by the SH$0$ES ($H_0^\text{R22} = 73.30 \pm 1.04$ km s$^{-1}$ Mpc$^{-1}$ \cite{Riess:2021jrx}) and Planck ($H_0^\text{P18} = 67.4 \pm 0.5$ km s$^{-1}$ Mpc$^{-1}$ \cite{Aghanim:2018eyx}) collaborations. {Note we opted for the slightly higher distance--ladder $H_0$ value \cite{Riess:2021jrx} which takes into account high redshift samples since this is better suited to test the robustness of the ABC \cite{Bernardo:2022pyz}.} We take these values as priors in the parameter space of $\Lambda$CDM, thus effectively organizing two models for the ABC to differentiate.

\item Redshift space distortions (RSD). RSD enables measurements of anisotropies in a galaxy distribution often due to the differences between galaxies peculiar velocities and the Hubble flow. This contains vital information about the growth of structure and consequently a blueprint of the formation and evolution of the perturbation, $\delta(z)$. This data is reported in terms of $f \sigma_8(z)$, which is a product of the growth rate $f(z) = d \ln \delta (z)/d \ln a(z)$ and the matter power spectrum $\sigma_8$ on scales of $8 h^{-1} {\rm Mpc}$. In standard cosmology, the matter density perturbation can be expressed analytically as \cite{Nesseris:2015fqa}
\begin{equation}
\label{eq:matterpert_lcdm}
    \delta(a) = a \,_2F_1\left( \dfrac{1}{3},1, \dfrac{11}{6}; a^3 \left(1 - \Omega_{m0}^{-1}\right)  \right) \,,
\end{equation}
{where $\,_2F_1\left(a,b,c;x\right)$ is the hypergeometric function.} For a variable dark energy equation of state, a more general analytical expression is also available \cite{Nesseris:2015fqa}. With this, the growth rate can be expressed as \cite{LeviSaid:2021yat}
\begin{equation}
\label{eq:growth_lcdm}
    f \sigma_8(z) = - \sigma_{8} (1 + z) \dfrac{\delta'(z)}{\delta(z = 0)} \,.
\end{equation}
We compare this prediction with the RSD measurements compiled in \cite{Kazantzidis:2018rnb} by considering the distance function
\begin{equation}
\begin{split}
    \chi^2_{\rm RSD} = \sum_{z_{\rm RSD}}
    & \left( f \sigma_8(z_{\rm RSD}) - f \sigma_{8, {\rm RSD}}(z_{\rm RSD}) \right) \\
    & \phantom{g} \times C^{-1}_{\rm RSD} \left( f \sigma_8(z_{\rm RSD}) - f \sigma_{8, {\rm RSD}}(z_{\rm RSD}) \right) \,,
\end{split}
\end{equation}
where the subscripts RSD stand for the observed values, and $C_{\rm RSD}$ is the covariance matrix.
\end{itemize}

{
We note that marginalization over $M$ with Pantheon$+$ data is now optional as the distance--moduli $\mu = m - M$ for each SNe have been provided (relative to their Cepheid hosts). However, for our work, we use instead the SNe apparent magnitudes, $m$, which remain accessible with the public data, since we are after measurements that are agnostic to the calibration of the distance--ladder. Otherwise, we would effectively be marginalizing twice over $M$ when considering $H_0$ priors since $M$ and $H_0$ are degenerate in SNe cosmology. In this way, the Hubble constants are given a fair treatment within the statistical framework of ABC.
}

{In addition, we bear in mind that the BAO $H(z)$ data hail from $\Lambda$CDM. The wiggle room our (and similar others such as \cite{Briffa:2020qli, Bonilla:2020wbn, Elizalde:2022rss, Mu:2023zct}) analyses play with is the uncertainty of these points that embody departures from model dependence, e.g., \cite{Bengaly:2021wgc} shows that cosmological reconstruction with the BAO predict \textit{ironically} deviations from $\Lambda$CDM and \cite{Bonilla:2020wbn} shows that BAO when used with CC and SNe can recreate local cosmological physics that is unfeeling to the exact value of the sound horizon radius. We also confirm that our main results are robust to the consideration of BAO.}

In the next section, we use MCMC and ABC together with three splices of the full data [(i) SNe $+$ CC (ii) SNe $+$ CC $+$ BAO (iii) SNe $+$ CC $+$ BAO $+$ RSD] to see how the results are affected, if it does, when considering the differing measurements.

\section{Results}
\label{sec:results}

We start by recollecting the two models we intend to compare and briefly discuss the MCMC results in Section \ref{subsec:mcmc}. We present the results of the ABC with the full and subsets of the data in Sections \ref{subsec:sne_cc}, \ref{subsec:sne_cc_bao}, and \ref{subsec:sne_cc_bao_rsd}.

\subsection{MCMC: P18 and R22}
\label{subsec:mcmc}

We emphasize that we consider only one cosmological model ($\Lambda$CDM) throughout. From this, we build two competing models by taking into account the Planck and SH$0$ES measurements of the Hubble constant as independent priors, while the rest of the cosmological parameters take initial flat priors. We refer to the models as P18 (with the prior $H_0^{\rm P18}$) and R22 (with the prior $H_0^{\rm R22}$) for brevity.

\begin{figure}
    \centering
    \includegraphics[width = 0.5 \textwidth]{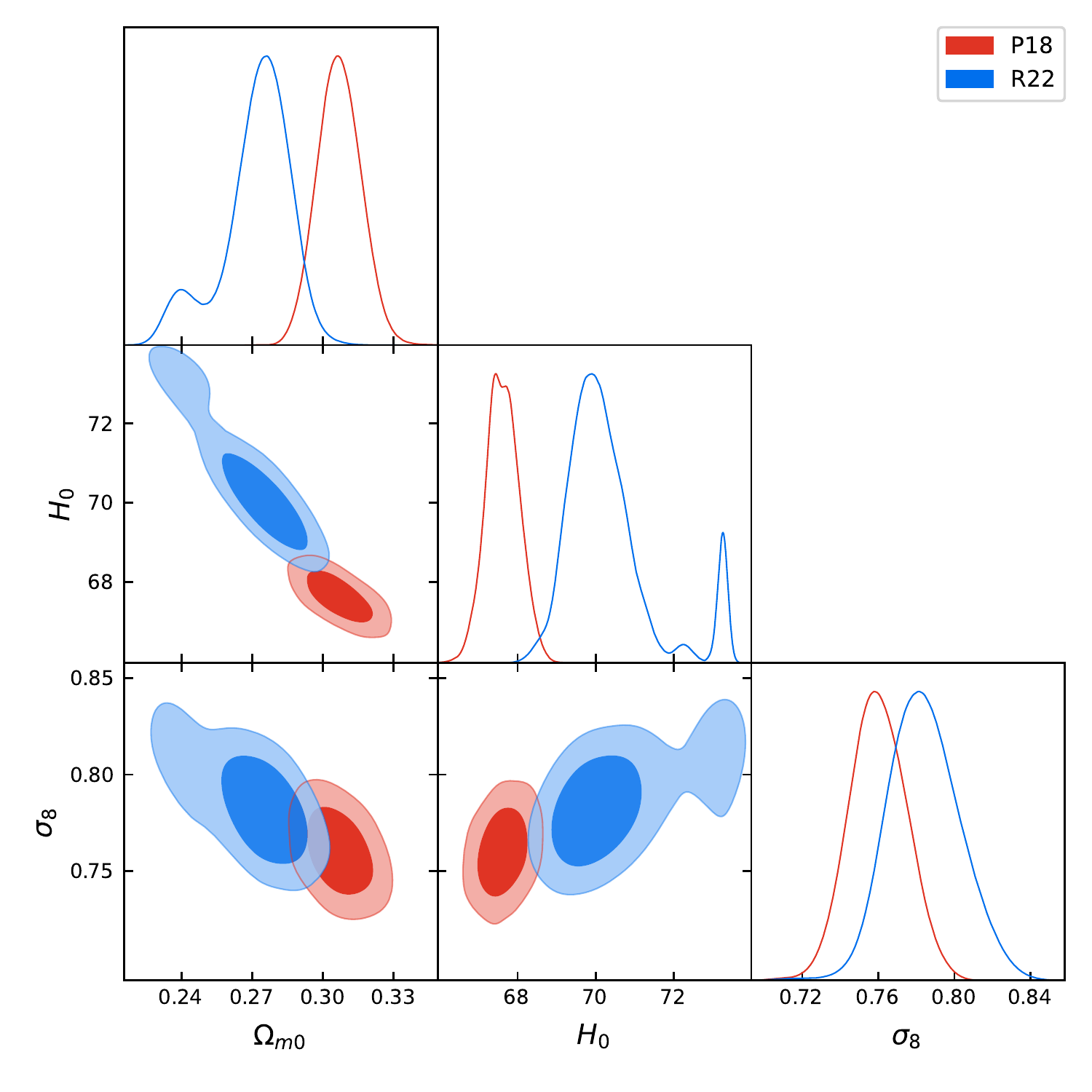}
    \caption{MCMC constraints on the $\Lambda$CDM cosmological parameters ($\Omega_{m0}, H_0, \sigma_8$) with the Hubble constant priors from the SH$0$ES ($H_0^{\rm R22} = 73.30 \pm 1.04$ km s$^{-1}$ Mpc$^{-1}$ \cite{Riess:2021jrx}) and Planck ($H_0^{\rm P18} = 67.4 \pm 0.5$ km s$^{-1}$ Mpc$^{-1}$ \cite{Aghanim:2018eyx}) teams.}
    \label{fig:mcmctriangle}
\end{figure}

To set a baseline for the ABC, we first look at the same models and data sets from the perspective of the traditional MCMC analysis. This brings us to the posterior shown in Figure \ref{fig:mcmctriangle} obtained using the full (background and perturbation) data set and Table \ref{tab:mcmc} which shows the constraints anchored on the full and subsets of the data. These results were obtained with a reasonable Gelman-Rubin convergence criterion of $R - 1 = 10^{-2}$.

Figure \ref{fig:mcmctriangle} explicitly shows the prior's influences on the background parameters ($H_0$ and $\Omega_{m0}$). This is pronounced for the R22 model, which takes in a high $H_0 \sim 73$ km s$^{-1}$Mpc$^{-1}$ prior, which is reflected in its posterior giving a slightly bimodal structure with a peak at $H_0 \sim 70$ km s$^{-1}$Mpc$^{-1}$ and a false maximum at $H_0 \sim 73$ km s$^{-1}$Mpc$^{-1}$. This understandably also translates to a low $\Omega_{m0} \sim 0.24$ peak in addition to the more reasonable matter density $\Omega_{m0} \sim 0.27$ since $H_0$ and $\Omega_{m0}$ are anti correlated through the high redshift CC points, $H(z)^2 \propto \Omega_{m0} h^2$ where $h = H_0/(100 \ {\rm km \ s}^{-1}{\rm Mpc}^{-1})$ \cite{Moresco:2016mzx}. A closer look at the P18 posterior on the Hubble constant also shows this, except in this case it is not too transparent as the prior turns out close to the estimate. It is worth noticing that the Hubble constant priors representative of the Hubble tension do not reflect as an independent tension in the matter power spectrum parameter $\sigma_8$, or in other words the $\sigma_8$ estimates significantly overlap (Figure \ref{fig:mcmctriangle}). This makes sense considering that the growth data fixes $\sigma_8$ as well as the matter density $\Omega_{m0}$ while the expansion history pins down $H_0$ and $\Omega_{m0}$. In Table \ref{tab:mcmc} we also take a look at the cosmological constraints when using MCMC and subsets of the full data. 

\begin{table}[h!]
    \centering
    \caption{Cosmological constraints on the $\Lambda$CDM model given Hubble constant priors by the Planck (P18) and the SH$0$ES (R22) values, obtained via the traditional MCMC analysis. The reduced chi-squared $\overline{\chi}^2 = \chi^2/N_{\rm data}$ where $N_{\rm data}$ is the number of data points considered in the sampling.}
    \begin{tabular}{|c|c|c|c|c|c|}
    \hline
    Data set & Model & $H_0$ [km s$^{-1}$Mpc$^{-1}$] & $\Omega_{m0}$ & $\sigma_8$ & {$\overline{\chi}^2$} \\ \hline \hline
    \multirow{2}{*}{SNe $+$ CC} &
    P18 & $67.4 \pm 0.5$ & $0.33 \pm 0.02$ & $-$ & {$0.8792$} \\ 
    & R22 & $72.1^{+1.1}_{-0.9}$ & $0.31 \pm 0.02$ & $-$ & {$0.8803$} \\ \hline
    \multirow{2}{*}{SNe $+$ CC $+$ BAO} &
    P18 & $67.6 \pm 0.4$ & $0.31 \pm 0.01$ & $-$ & {$0.8836$} \\ 
    & R22 & $70.3^{+0.4}_{-1.0}$ & $0.27^{+0.02}_{-0.01}$ & $-$ & {$0.8839$} \\ \hline
    \multirow{2}{*}{\phantom{g} SNe $+$ CC $+$ BAO $+$ RSD \phantom{g}} &
    P18 & $67.6^{+0.4}_{-0.5}$ & $0.31 \pm 0.01$ & $0.75 ^{+0.02}_{-0.01}$ & {$0.8704$} \\ 
    & R22 & $70.7 ^{+2.5}_{-1.5}$ & $0.27^{+0.02}_{-0.01} $ & $0.78^{+0.03}_{-0.01}$ & {$0.8720$} \\ \hline
    \end{tabular}
    \label{tab:mcmc}
\end{table}

An immediate observation is that without the BAO the matter density constraints turn out peculiarly high with P18. This is nonetheless consistent with the Planck value at $68\%$ confidence. The addition of the BAO can be seen to have made the matter sector more consistent with P18 \cite{Aghanim:2018eyx, DiValentino:2019ffd}. In terms of the $H_0$, it is clear that the BAO compels the Hubble constant to a lower value, consistent with other results. This is the case with both P18 and R22 models. However, with the BAO, we find the matter fraction estimate in R22 to be quite low, but notably still consistent with the Planck value at $95\%$ confidence. The inclusion of the RSD growth observations, adding cosmological perturbations in the mix, now better constrains the matter sector \cite{Kazantzidis:2018rnb}. Interestingly, we find the matter sector constraints to be consistent with and without the RSD as long as the BAO is considered. Having fixed the matter fraction, the RSD now is able to further constrain one more parameter, that is, the smoothed matter perturbation amplitude, $\sigma_8$. We find reasonable results for the matter sector including the perturbation. The Hubble constant value also remained consistent with the Planck measurement regardless of the prior used.

We find Hubble constants consistent with the Planck value when considering the expansion rate data with the BAO. Without the BAO, as a matter of fact, only the R22 model produces an $H_0$ that is in tension with the Planck estimate. This may be because the background data is inconsistent with the larger $H_0$ prior that it considers. Regardless, the addition of the BAO lowers $H_0$ while we see that the RSD fixes the matter sector in addition to the perturbations.

We lastly comment on which model, given a data set, MCMC prefers. This can be assessed using various metrics, with the Bayesian evidence being the more rigorous one. However, since we are dealing with only models with the same number of parameters {and that our likelihoods dominate over the prior in all of our cases}, we may simply resort to a comparison of the likelihoods, ${\cal L}$, or the chi-squared statistic,  $\chi^2 \propto -2 \log {\cal L}$ as a measure of the goodness of fit \cite{Trotta:2008qt}. With this in mind, we find that in all cases (Table \ref{tab:mcmc}), the P18 model produces a better fit than R22. This is intriguing in light of the Hubble tension which sees a significant statistical deviation between local cosmological physics and the cosmic microwave background. We find this is supported by the ABC, this work's main theme, as we proceed to show.

\subsection{ABC: SNe + CC}
\label{subsec:sne_cc}

\begin{figure}
    \centering
	\subfigure[]{
		\includegraphics[width = 0.49 \textwidth]{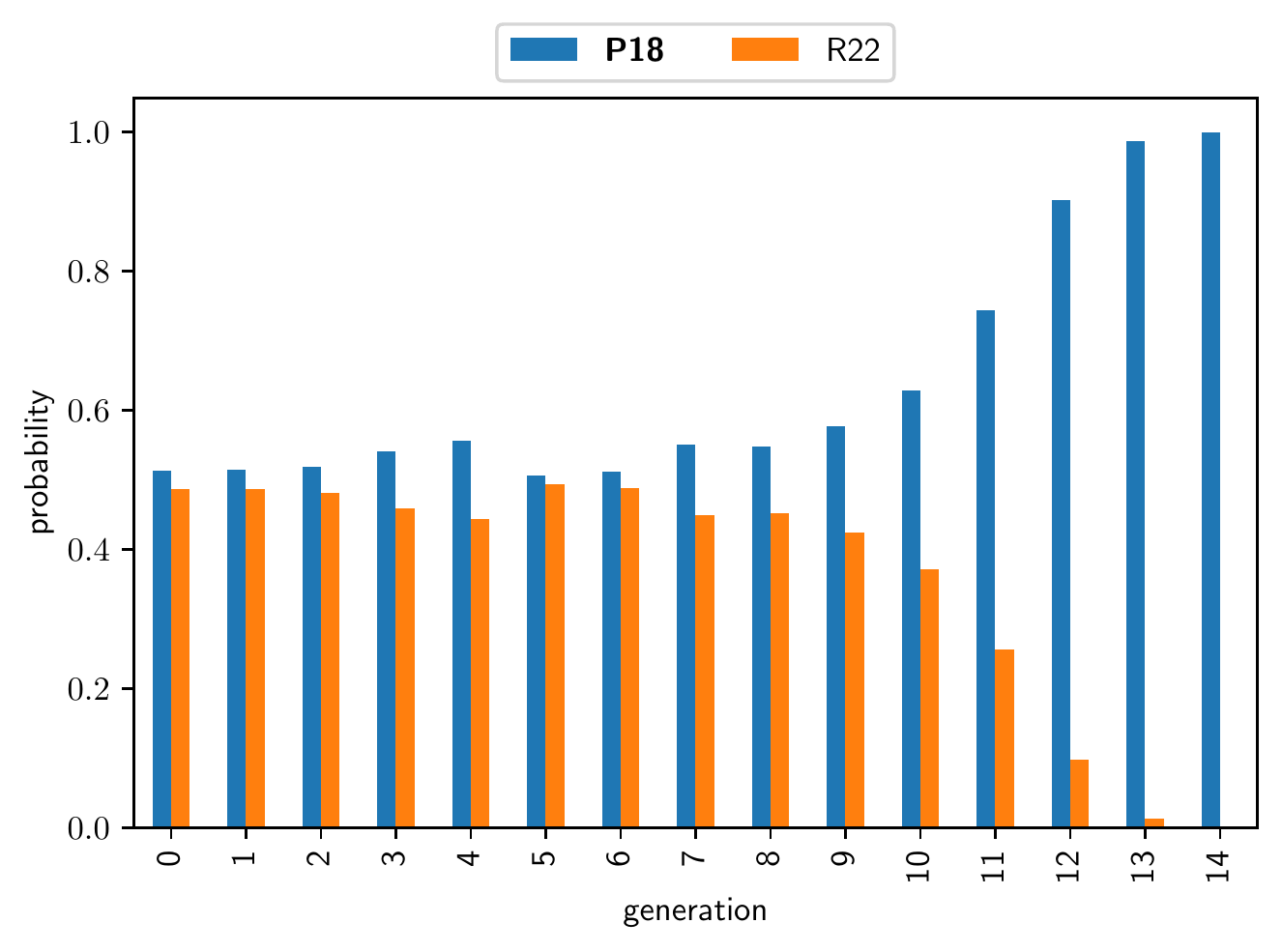}}
    \subfigure[]{
		\includegraphics[width = 0.49 \textwidth]{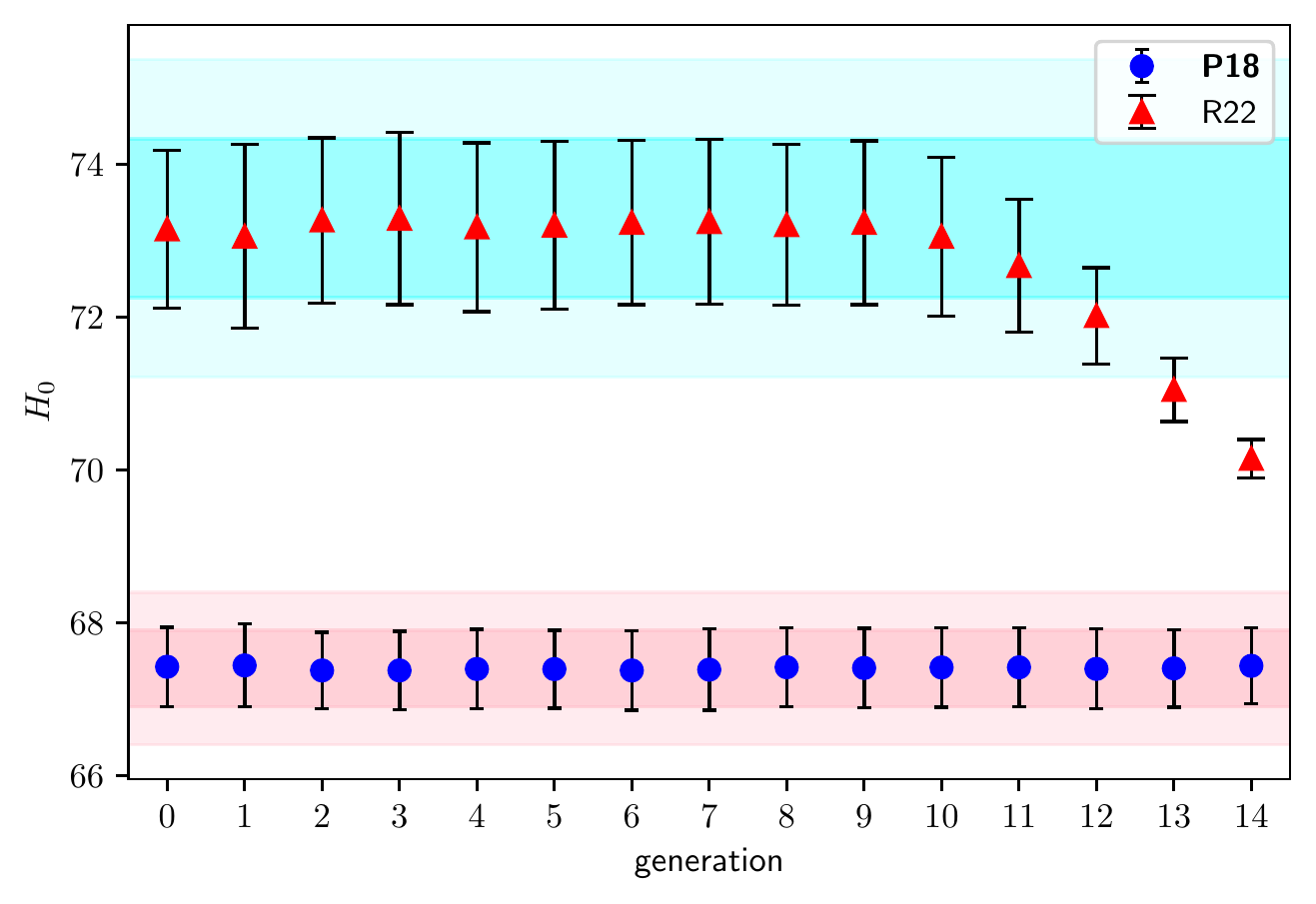}}
    \caption{(a) Model posterior ABC evolution with SNe and CC and (b) the generational Hubble constant estimates. The pink and blue bands in (b) correspond to $68\%$ and $95\%$ confidence limits of the Planck ($H_0^\text{P18} = 67.4 \pm 0.5$ km s$^{-1}$Mpc$^{-1}$ \cite{Riess:2021jrx}) and SH$0$ES ($H_0^\text{R22} = 73.30 \pm 1.04$ km s$^{-1}$Mpc$^{-1}$ \cite{Aghanim:2018eyx}) estimates of the Hubble constant.}
    \label{fig:modelpost_SNeCC}
\end{figure}

The base of our data set consists of brightness observations of supernovae and expansion rate measurements from cosmic chronometers. Both provides indirect and direct independent observations of the cosmological background. Figure \ref{fig:modelpost_SNeCC} shows the evolution of the model posterior and the Hubble constant in each model per generation.

We find here that the model taking in the Planck Hubble constant as a prior overtakes the one considering the SH$0$ES estimate. The same result can be obtained with varying size of the population in each generation. Figure \ref{fig:modelpost_SNeCC}(a) shows that the P18 model takes over the competition by the $6$th generation and completely dominates by the $13$th generation, when the R22 model's population becomes extinct. To understand what occurred during the course of the evolution, we take a look at the generational evolution of the Hubble constant in Figure \ref{fig:modelpost_SNeCC}(b). This shows that as the selection becomes more competitive, as the ABC narrows down the acceptable distance function values, the Hubble constant estimate in the R22 model falls outside of its prior to be able to keep up; however, at the same time, going away from its prior increases the overall distance and so its population shrinks as the Hubble constant falls down. We can see this happened more clearly after the $9$th generation in Figure \ref{fig:modelpost_SNeCC}(b), which is after the $6$th generation when the P18 model has taken control of the competition.

The takeaway from this is that as far as ABC is concerned the Planck Hubble constant estimate is more consistent with the late time observations (SNe + CC) than with the SH$0$ES value. This could be taken as a surprising result considering that we are basing this on late time observations. However, we are about to find out that this is robust even if we consider other data sets.

We note that as a model selection tool only the parameters estimated in the surviving model after the ABC run are meaningful. In this case, we get the $H_0$ estimate consistent with the Planck data, as also illustrated in Figure \ref{fig:modelpost_SNeCC} where the P18 generational $H_0$ values coincide with the confidence contour provided by the Planck measurement.

\subsection{ABC: SNe + CC + BAO}
\label{subsec:sne_cc_bao}

Now we consider expansion rate measurements from BAO. These additional measurements can be considered more stringent as the error bars are smaller. However, we keep in mind that the BAO depend on a prescription of the physics during the Hubble drag through the sound horizon radius. We present our results with this more constraining data combination, i.e., $H_{\rm BAO}\left(z_{\rm BAO}\right)$ have relatively lower error bars compared with CC measurements.

\begin{figure}
    \centering
	\subfigure[]{
		\includegraphics[width = 0.49 \textwidth]{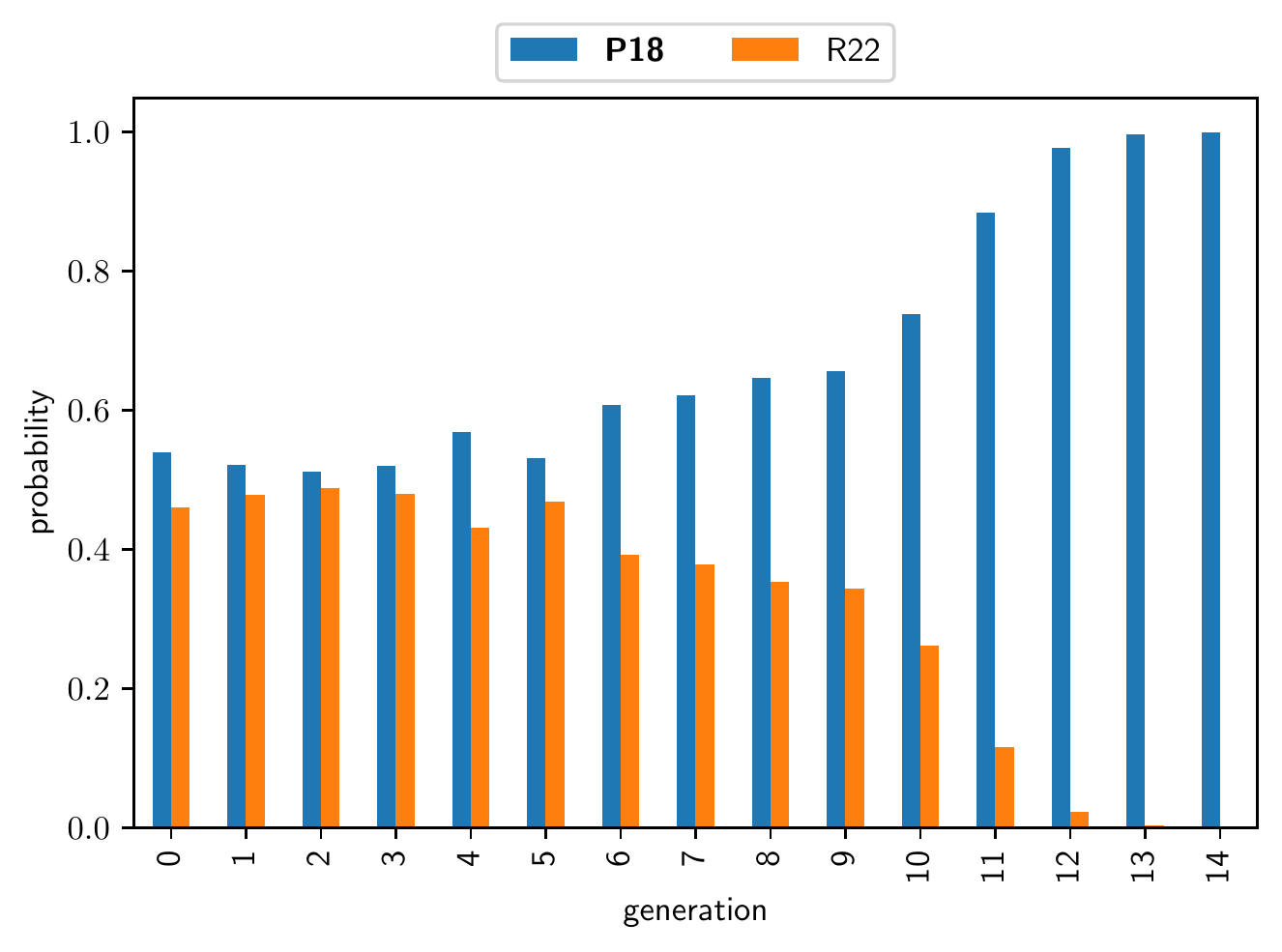}}
    \subfigure[]{
		\includegraphics[width = 0.49 \textwidth]{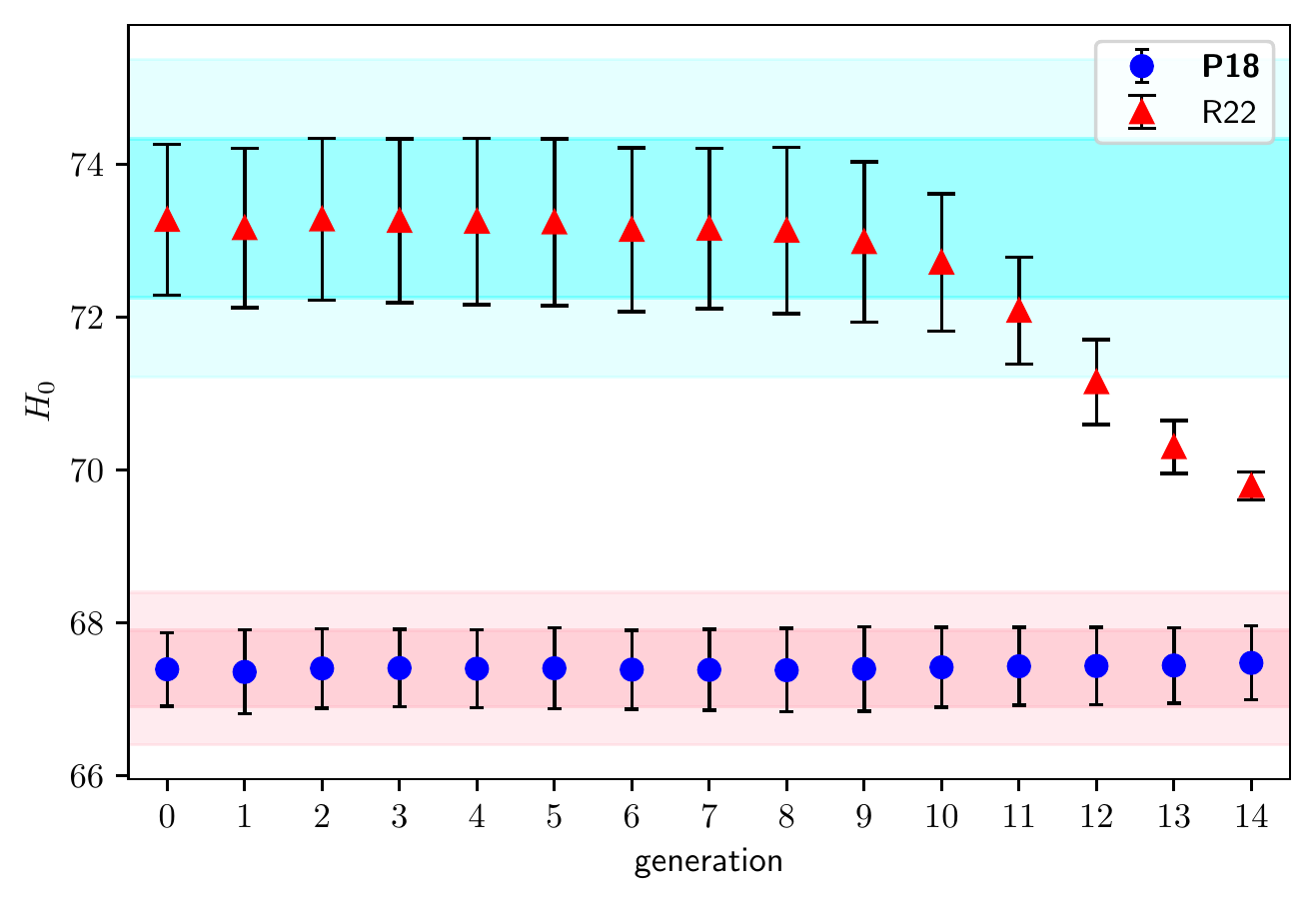}}
    \caption{(a) Model posterior ABC evolution with SNe, CC, and BAO compiled data sets and (b) the generational Hubble constant estimates. The pink and blue bands in (b) correspond to $68\%$ and $95\%$ confidence limits of the Planck ($H_0^\text{P18} = 67.4 \pm 0.5$ km s$^{-1}$Mpc$^{-1}$ \cite{Riess:2021jrx}) and SH$0$ES ($H_0^\text{R22} = 73.30 \pm 1.04$ km s$^{-1}$Mpc$^{-1}$ \cite{Aghanim:2018eyx}) estimates of the Hubble constant.}
    \label{fig:modelpost_SNeCCBAO}
\end{figure}

Perhaps the most interesting thing observation here is that nearly the same picture emerged with the addition of the BAO. It took the same number of generations for the ABC to come out with one model, P18, and that the natural selection became one sided after about six generations. This is clearly shown in Figure \ref{fig:modelpost_SNeCCBAO}, which if one looks away may appear just like Figure \ref{fig:modelpost_SNeCC} without the BAO. This strengthens the earlier result of the Planck $H_0$ being statistically preferred by ABC and the data. We see that the Hubble constant estimates of P18 also remained consistent throughout the evolution, all the while the R22 values drop to keep up as the competition tightens, but eventually loses its population.

Thus we see that even with BAO the ABC shows that the Planck Hubble constant estimate \textit{as a prior} is more consistent with observations. Understandably we have to be skeptical about this as the Planck value is measured from the cosmic microwave background rather than local physics. But next we also take a look at the perturbations to see if this influences the result.

\subsection{ABC: SNe + CC + BAO + RSD}
\label{subsec:sne_cc_bao_rsd}

\begin{figure}
    \centering
	\subfigure[]{
		\includegraphics[width = 0.49 \textwidth]{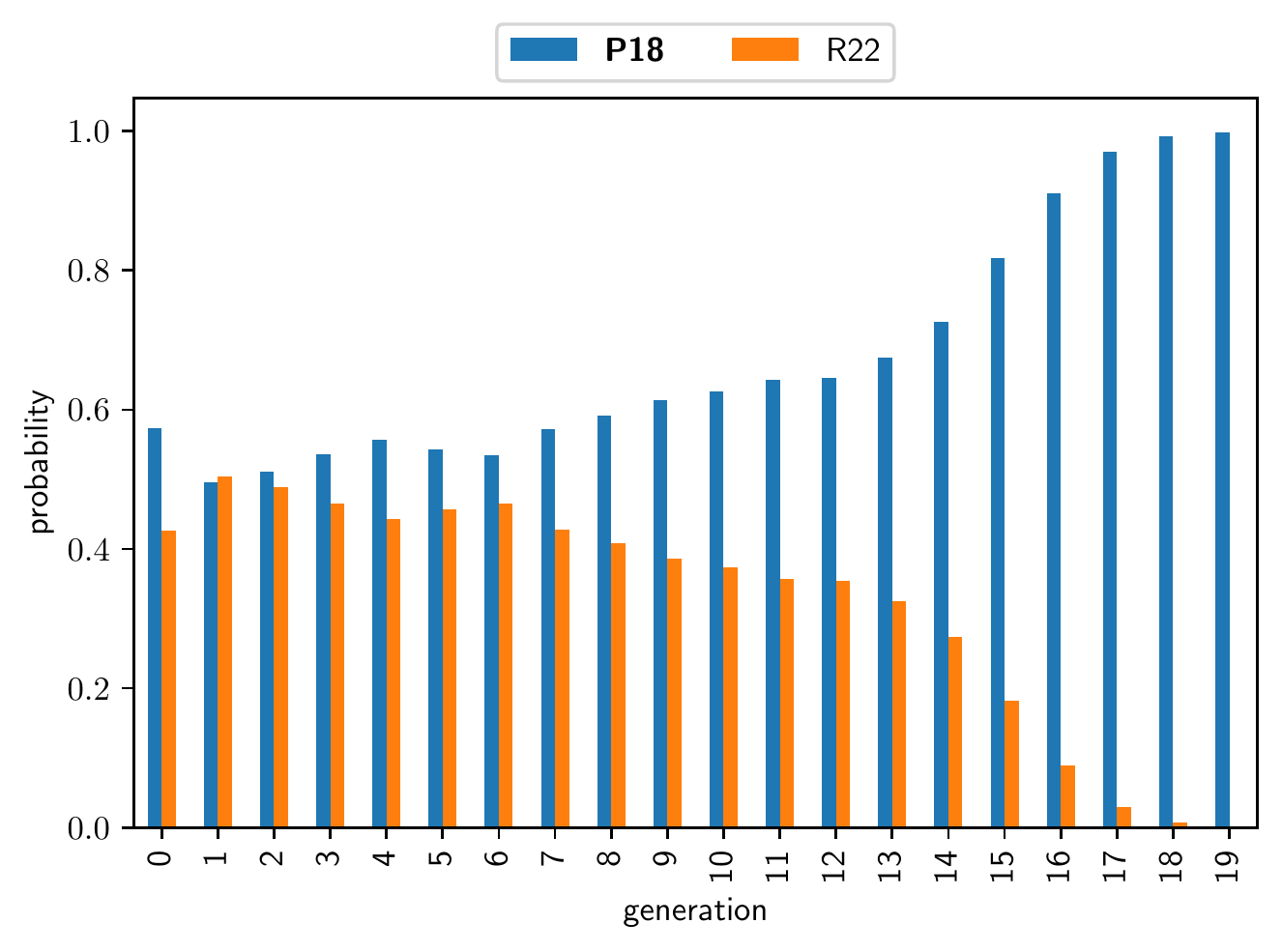}}
    \subfigure[]{
		\includegraphics[width = 0.49 \textwidth]{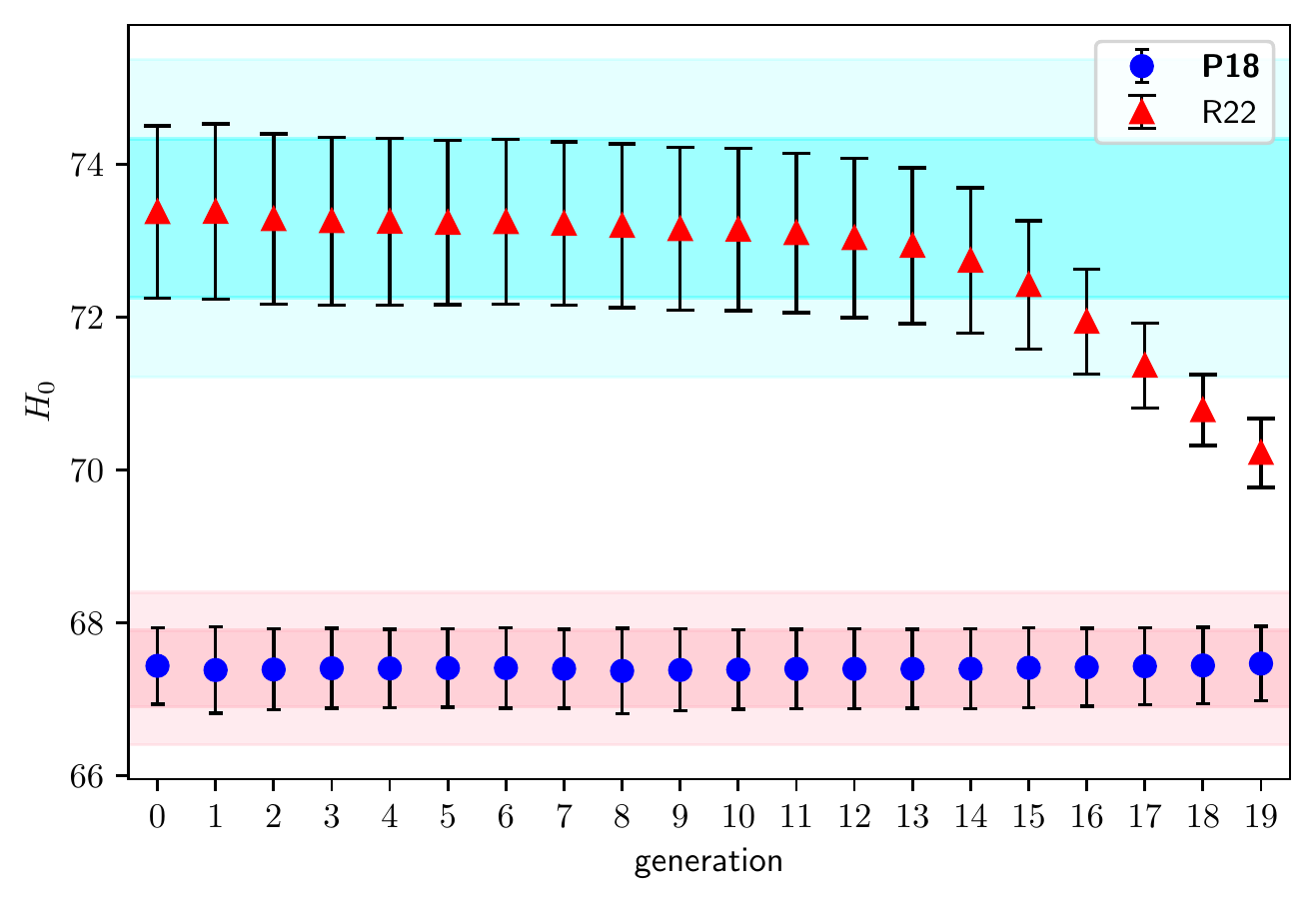}}
    \caption{(a) Model posterior ABC evolution with SNe, CC, BAO, and RSD background and perturbation data and (b) the generational Hubble constant estimates. The pink and blue bands in (b) correspond to $68\%$ and $95\%$ confidence limits of the Planck ($H_0^\text{P18} = 67.4 \pm 0.5$ km s$^{-1}$Mpc$^{-1}$ \cite{Riess:2021jrx}) and SH$0$ES ($H_0^\text{R22} = 73.30 \pm 1.04$ km s$^{-1}$Mpc$^{-1}$ \cite{Aghanim:2018eyx}) estimates of the Hubble constant.}
    \label{fig:modelpost_SNeCCBAORSD}
\end{figure}

We consider this time not only background cosmological data (SNe $+$ CC $+$ BAO) but also the leading order, linear perturbations of the matter density through RSD. The results now take the input from the structure formation. The relevant ABC results are shown in Figure \ref{fig:modelpost_SNeCCBAORSD}.

We find that the addition of the growth data has made the selection drag for a longer period, which we may expect since the algorithm needs extra time to compare the simulations with bigger data. This time, it took $18$ generations before the P18 model completely overcome the R22 model. After this, the cosmological parameters in P18 were being estimated just like in MCMC. The main takeaway is that even with the growth data constraining the matter sector, the $\Lambda$CDM model anchored on the Planck Hubble constant turns out to be more consistent with the observations.

Looking at the Hubble constant estimates per generation, we see that eventually the R22 model slips out of the competition when it has to compromise with an $H_0$ that is away from its prior. Eventually to be able to keep up with P18 its Hubble constant estimate deviates too far outside the SH$0$ES value (R22's $H_0$ prior), making the overall distance measurements of the members of its population too large. This shrinks the population's size, leading to its collapse. All the while, it can be seen that the Hubble constant estimate in the P18 model remained convincingly consistent throughout, thereby stifling its competitor.

\begin{figure}
    \centering
    \includegraphics[width = 0.5\textwidth]{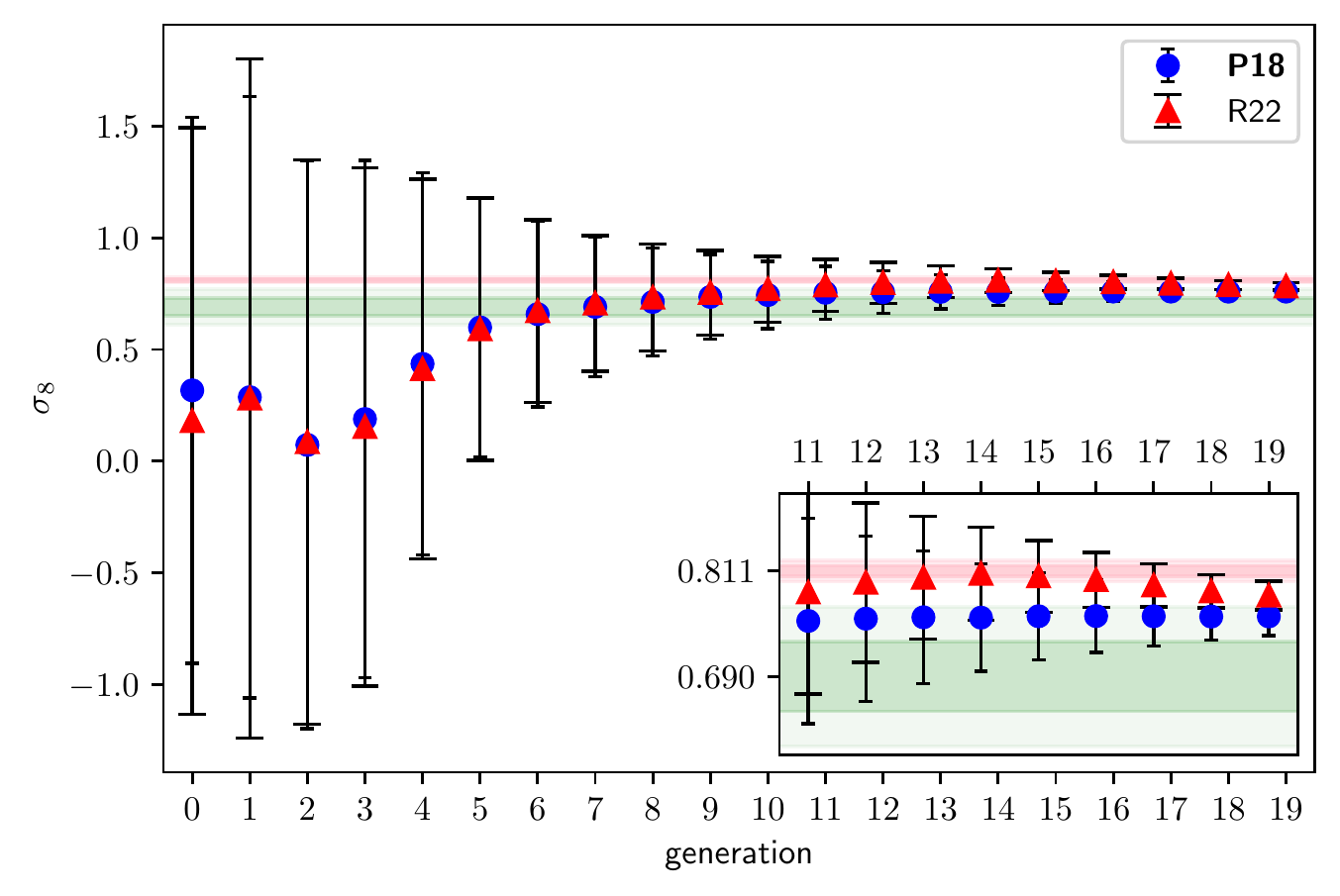}
    \caption{Generational estimates of the smoothed amplitude of the matter power spectrum, $\sigma_8$, given the SNe $+$ CC $+$ BAO $+$ RSD data set. The inset shows a zoomed-in view beginning at generation $11$. The pink and green bands show the $68\%$ and $95\%$ confidence limits of the corresponding cosmic microwave background ($\sigma_8 = 0.811 \pm 0.006$ \cite{Aghanim:2018eyx}) and galaxy clustering ($\sigma_8 = 0.69 \pm 0.04$ \cite{Philcox:2021kcw}) reference values.}
    \label{fig:sigma8ABC}
\end{figure}

Since we are considering growth data, it is beneficial to also look at the generational estimates of the physical quantity $\sigma_8$. We present the ABC evolution in Figure \ref{fig:sigma8ABC}.

The large initial error bars reflect the initial flat priors on the matter sector. The evolution eventually gets to reasonable values as it populations compete for their survival. Interestingly, the results show $\sigma_8$ estimates that are consistent with galaxy clustering \cite{Philcox:2021kcw} (green band in Figure \ref{fig:sigma8ABC}). However, we take this with a grain of salt as the ABC estimates fall toward to edge of the galaxy clustering confidence interval. This nonetheless puts ABC's view of the $\sigma_8$ tension given the data sets we consider here.

{Now, before we leave the section, we point out a possible misinterpretation that the ABC is not learning anything from the data since the P18 model's $H_0$ estimates via Figures \ref{fig:modelpost_SNeCC}--\ref{fig:modelpost_SNeCCBAORSD}(b) stay about the same. This is however not true as the models' posterior evolution embodied by \ref{fig:modelpost_SNeCC}--\ref{fig:modelpost_SNeCCBAORSD}(a) are reflective of the learning process. The explanation is simply: in the ABC, the generational estimates -- their model probabilities and parameters -- are not separate but are part of one process, that is aimed at singling out the more competitive model. The learning process can only be viewed consistently from the whole, e.g., P18 \textit{and} R22 for this work. Regardless, in the next section, we play with the $H_0$ priors, thus no longer working with the P18 and R22 models, to further showcase the ABC-SMC's robustness.}

\section{{Varying $H_0$ priors}}
\label{sec:varyingH0priors}

{
To elucidate more on the robustness of the ABC when it comes to the dependence on the prior, we test its results with the full data set, but instead of considering the Planck and SH$0$ES Gaussian $H_0$ priors, we take flat and wider initial $H_0$ priors, i.e., 
\begin{equation}{\cal N}\left( H_0, \sigma_{H_0} \right) \rightarrow {\cal U}\left( H_0 - 2 \sigma_{H_0}, H_0 + 2 \sigma_{H_0} \right) \,, \nonumber \end{equation}
where ${\cal N}$ and ${\cal U}$ are the normal and uniform distributions, respectively, and $H_0 \pm \sigma_{H_0}$ corresponds an $H_0$ estimate such as that of Planck and SH$0$ES. We also consider another extreme example of a very wide $H_0$ initial flat priors. The results of both after being fed into the ABC-SMC model selection algorithm are shown below in Figure \ref{fig:AlternateH0Priors}.
}

\begin{figure}[h!]
    \centering
	\subfigure[ Flat $H_0$ priors ]{
		\includegraphics[width = 0.48 \textwidth]{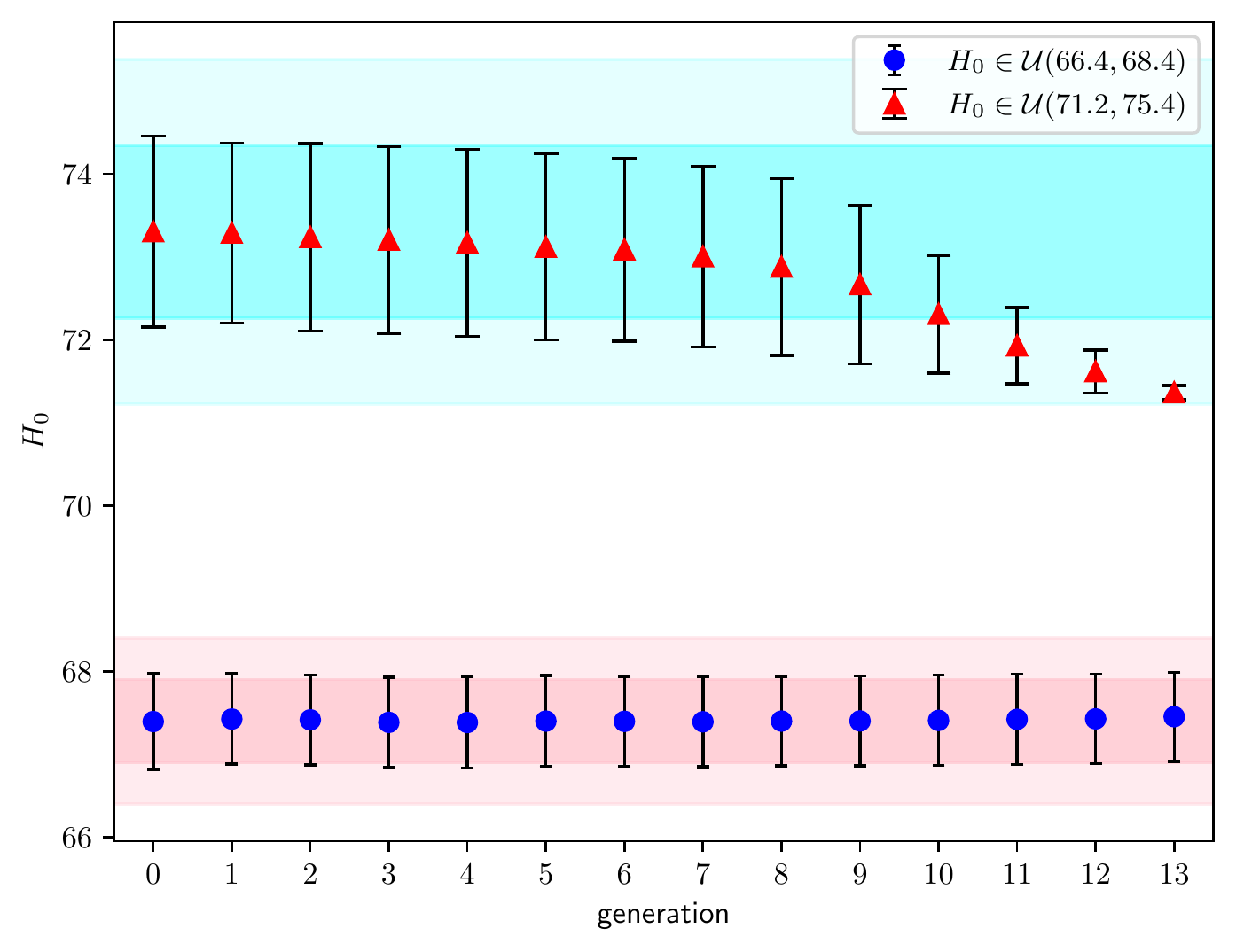}}
	\subfigure[ Flat $H_0$ priors ]{
		\includegraphics[width = 0.48 \textwidth]{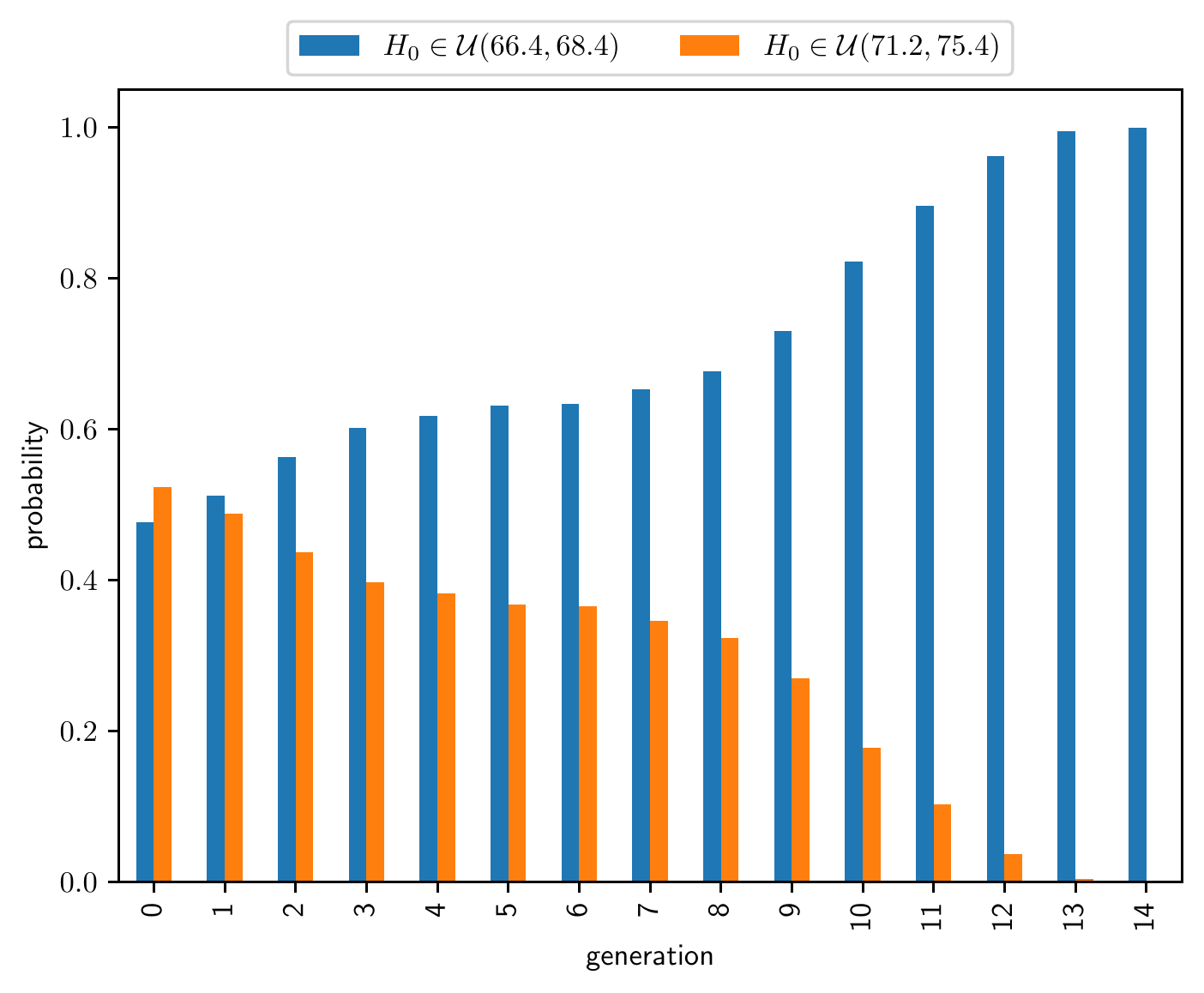}}
    \subfigure[Flat-Super Wide $H_0$ priors]{
		\includegraphics[width = 0.48 \textwidth]{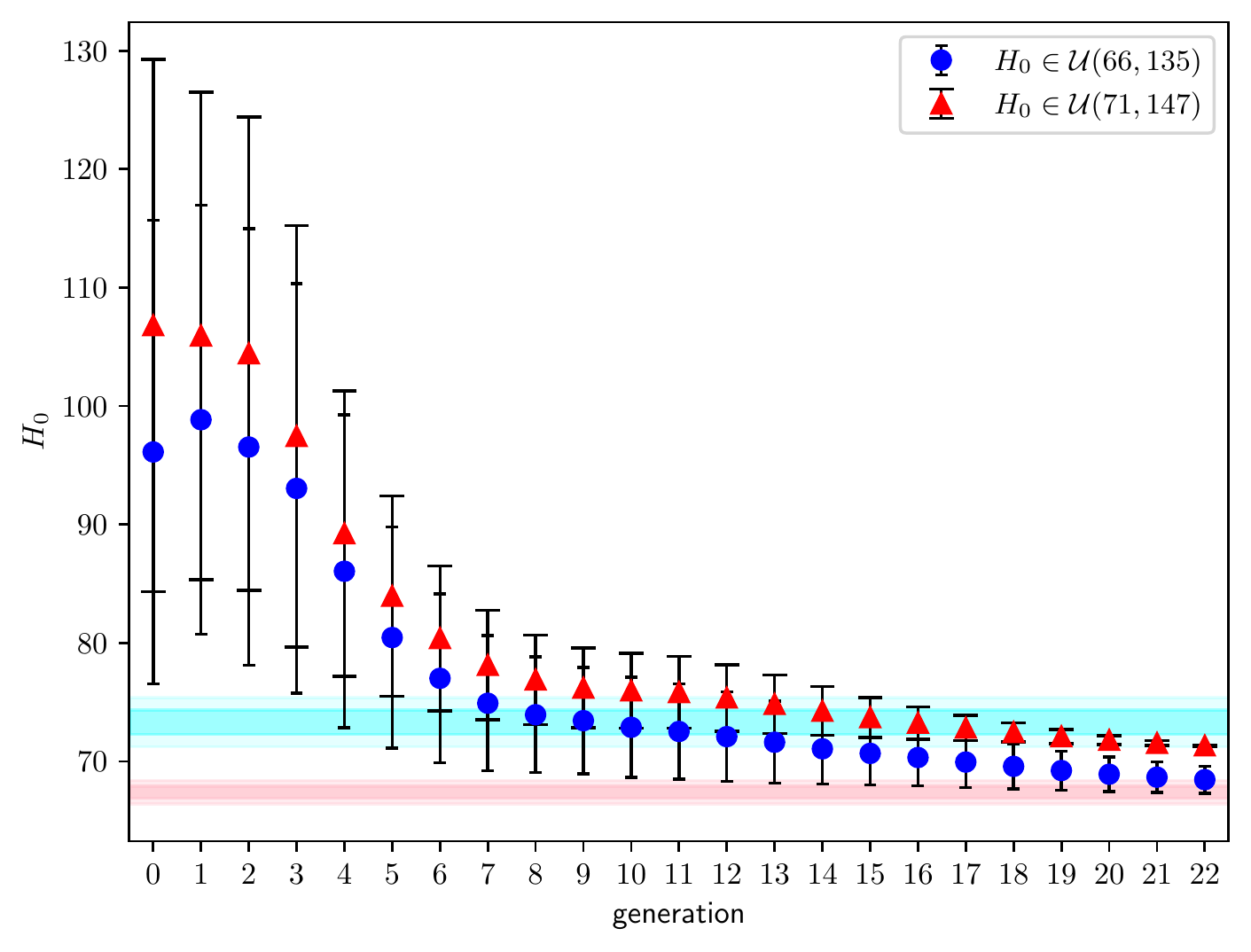}}
    \subfigure[Flat-Super Wide $H_0$ priors]{
		\includegraphics[width = 0.48 \textwidth]{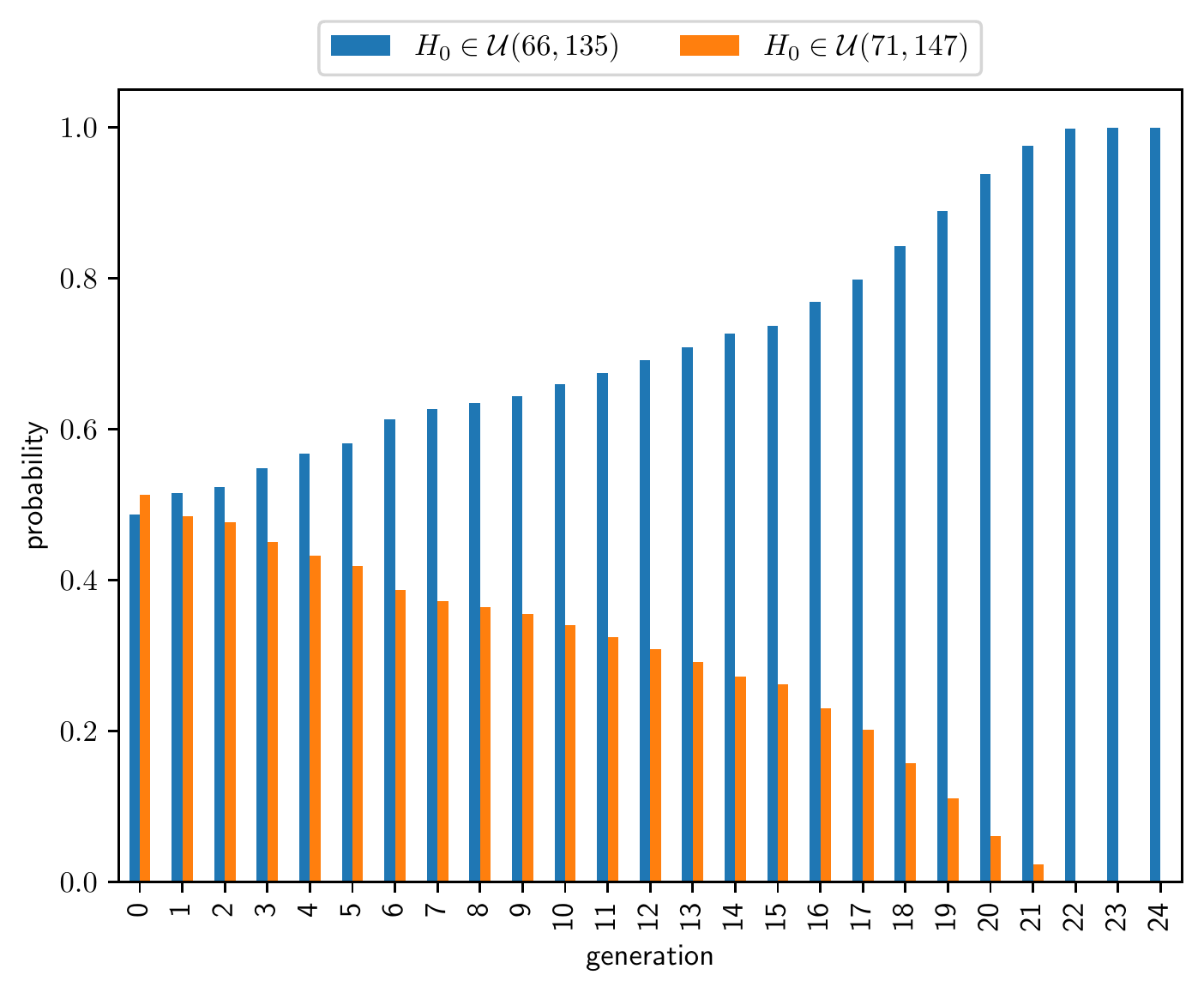}}
    \caption{{ABC generational $H_0$ evolution with flat initial priors $H_0 \in (H_0^{\rm min}, H_0^{\rm max})$ for the full data set SNe + CC + BAO + RSD.}} \label{fig:AlternateH0Priors}
\end{figure}

{
First off, we note that these are not the P18 and R22 models we have considered previously that are defined by the well-known CMB and distance--ladder estimates of $H_0$. Nonetheless, the message of Figure \ref{fig:AlternateH0Priors} is that the dependence of the ABC on the initial prior is irrelevant owing to its evolution algorithm. In both cases, the model with the lower $H_0$ values comes out naturally selected with an $H_0$ estimate consistent with the Planck best fit (pink band) compared with the SH$0$ES one (blue band). This feature is in stark contrast with MCMC which are strongly prior influenced. On the other hand, in the ABC, the parameter estimates are dependent on the evolution parameters such as the number of generations. This largely gives the pros and cons of the ABC and the MCMC with respect to each other. In fact the ABC-SMC pioneers \cite{2009arXiv0901.1925T} have teased to take the ABC model selection results, but then independently estimate the parameters of each model, say, with MCMC. This acknowledges the inherent limitations of both and thus take the best of both worlds. We are definitely heading to this promising direction in a future work as we continue to advocate the ABC as a model selection tool for cosmology.
}

\section{Conclusions}
\label{sec:conclusions}

{The summary of the constraints obtained this work is presented in Table \ref{tab:mcmcandabc}. We stress that in the ABC only the parameters of the deemed `naturally selected' model are meaningful. Nonetheless, the parameters of the extincted models based on its last surviving generation can be presented for illustrative purposes such as in Table \ref{tab:mcmcandabc}.}

\begin{table}[h!]
    \centering
    \caption{{Summary of $H_0$ and $\sigma_8$ results of MCMC and ABC based on the $\Lambda$CDM model and priors by the Planck (P18) and the SH$0$ES (R22) values. For the ABC, the labels and estimates of the `naturally selected' model are in \textbf{bold face}. The parameters of the collapsed model determined by its last population before extinction are included for comparison purposes only.}}
    \begin{tabular}{|c|c|c|c|}
        \hline
        {Stat. Method : Data set (data points)} & {Model} & {$H_0$ [km s$^{-1}$Mpc$^{-1}$]} & {$\sigma_8$} \\ \hline \hline
        \multirow{2}{*}{{MCMC : SNe $+$ CC (1621)}} &
        {P18} & {$67.4 \pm 0.5$} &  {$-$} \\ 
        & {R22} & {$72.1^{+1.1}_{-0.9}$} & {$-$} \\ \hline
        \multirow{2}{*}{{MCMC : SNe $+$ CC $+$ BAO (1647)}} &
        {P18} & {$67.6 \pm 0.4$} & {$-$} \\ 
        & {R22} & {$70.3^{+0.4}_{-1.0}$} & {$-$} \\ \hline
        \multirow{2}{*}{{MCMC : SNe $+$ CC $+$ BAO $+$ RSD (1710)}} &
        {P18} & {$67.6^{+0.4}_{-0.5}$} & {$0.75 ^{+0.02}_{-0.01}$} \\ 
        & {R22} & {$70.7 ^{+2.5}_{-1.5}$} & {$0.78^{+0.03}_{-0.01}$} \\ \hline
        \multirow{2}{*}{{\textbf{ABC} : SNe $+$ CC (1621)}} &
        \textbf{{P18}} & \textbf{{$\mathbf{67.4 \pm 0.5}$}} & {$-$} \\ 
        & {R22} & {$70.1 \pm 0.3$} & {$-$} \\ \hline
        \multirow{2}{*}{{\textbf{ABC} : SNe $+$ CC $+$ BAO (1647)}} &
        \textbf{{P18}} & \textbf{{$\mathbf{67.5 \pm 0.5}$}} & {$-$} \\ 
        & {R22} & {$69.8 \pm 0.2$} & {$-$} \\ \hline
        \multirow{2}{*}{{\textbf{ABC} : SNe $+$ CC $+$ BAO $+$ RSD (1710)}} &
        \textbf{{P18}} & \textbf{{$\mathbf{67.5 \pm 0.5}$}} & \textbf{{$\mathbf{0.76 \pm 0.02}$}} \\ 
        & {R22} & {$70.2 \pm 0.5$} & {$0.78 \pm 0.02$} \\ \hline
    \end{tabular}
    \label{tab:mcmcandabc}
\end{table}

We ought to mention that in this work the Hubble constants are compared through ABC, which is very much different with \cite{Bernardo:2022pyz} where dark energy models are juxtaposed by the algorithm. In other words, this work gives ABC's direct insight into the Hubble tension by pitting the Hubble constants against each other in a dynamic scheme. Our results furthermore show up as a testament to ABC's decisiveness and overall potential for model selection in cosmology in the forthcoming data--driven era. 

We find that the Approximate Bayesian Computation naturally favors the Planck value of the Hubble constant. This holds when considering both cosmological background and matter perturbation data. Our takeaway from this is that the late Universe estimate of the Hubble constant is consistent with that of Planck. Of course, we get this from a noncanonical approach and so must be viewed with skepticism. An interesting result is that the traditional MCMC analysis agrees with this, in the sense that the analysis based on the Planck Hubble constant prior lead to a better fit compared with the one coming from SH$0$ES. It is also worthwhile to point out that the ABC's preference for a low Hubble constant consistent with the Planck value goes beyond the scope of the standard cosmological model \cite{Bernardo:2022pyz}. This motivates the use of some other independent statistical methods for later cosmological analysis. The ABC is only one such approach that looks into cosmology with fresh eyes, but there are also others that may be more mainstream such as neural networks \cite{Grandon:2022gdr, Mukherjee:2022yyq} and genetic algorithms \cite{Bogdanos:2009ib, Alestas:2022gcg}. It will be worth the time looking into all these together.

We emphasize that we do not solve or propose a physics solution to the Hubble tension. Rather, the ABC brings into the table a fresh new insight that the Planck's Hubble constant is consistent with late time cosmology, contrary to the Hubble tension's premise. Granted, the reliability of the ABC in cosmological model selection and parameter estimation remains to be established further. One angle is it leads to perhaps unreasonably too narrow parameter estimates when the number of generations is taken too large \cite{Bernardo:2022pyz, 2009arXiv0901.1925T}. Model selection in ABC also does not take into account, at least superficially, the number of parameters of the models and therefore has no safeguard for overfitting. {The model parameters post-ABC-SMC could for example be obtained independently such as with MCMC to have perhaps more reasonable estimates, as teased in \cite{2009arXiv0901.1925T}. These shortcomings point to ABC-SMC as an unfinished product for cosmological application, but we have to regardless acknowledge the pros that it brings to the table such as its likelihood-free and prior insensitive features. An ironed out product taking the best of MCMC and ABC-SMC for model selection and parameter estimatation in cosmology is definitely a worthy future work.} For the meantime, it is always nice and worthwhile to try alternative statistical methods, like we do with cosmological models. An assessment tool that transcends different statistical approaches will be useful to establish in this regard. With openness to new methods, we may just find the Universe to be more exciting, which may bring together an adventure among various fields in solving physics problems.

\section*{Acknowledgements}
The initial stages of this work were done in part during the TCA Summer Student Program 2022 hosted by the National Center for Theoretical Sciences.

\appendix

\section{{Matter in ABC}}
\label{sec:matterABC}

\begin{figure}[h!]
    \centering
	\subfigure[ \ \ {SNe + CC} ]{
		\includegraphics[width = 0.5 \textwidth]{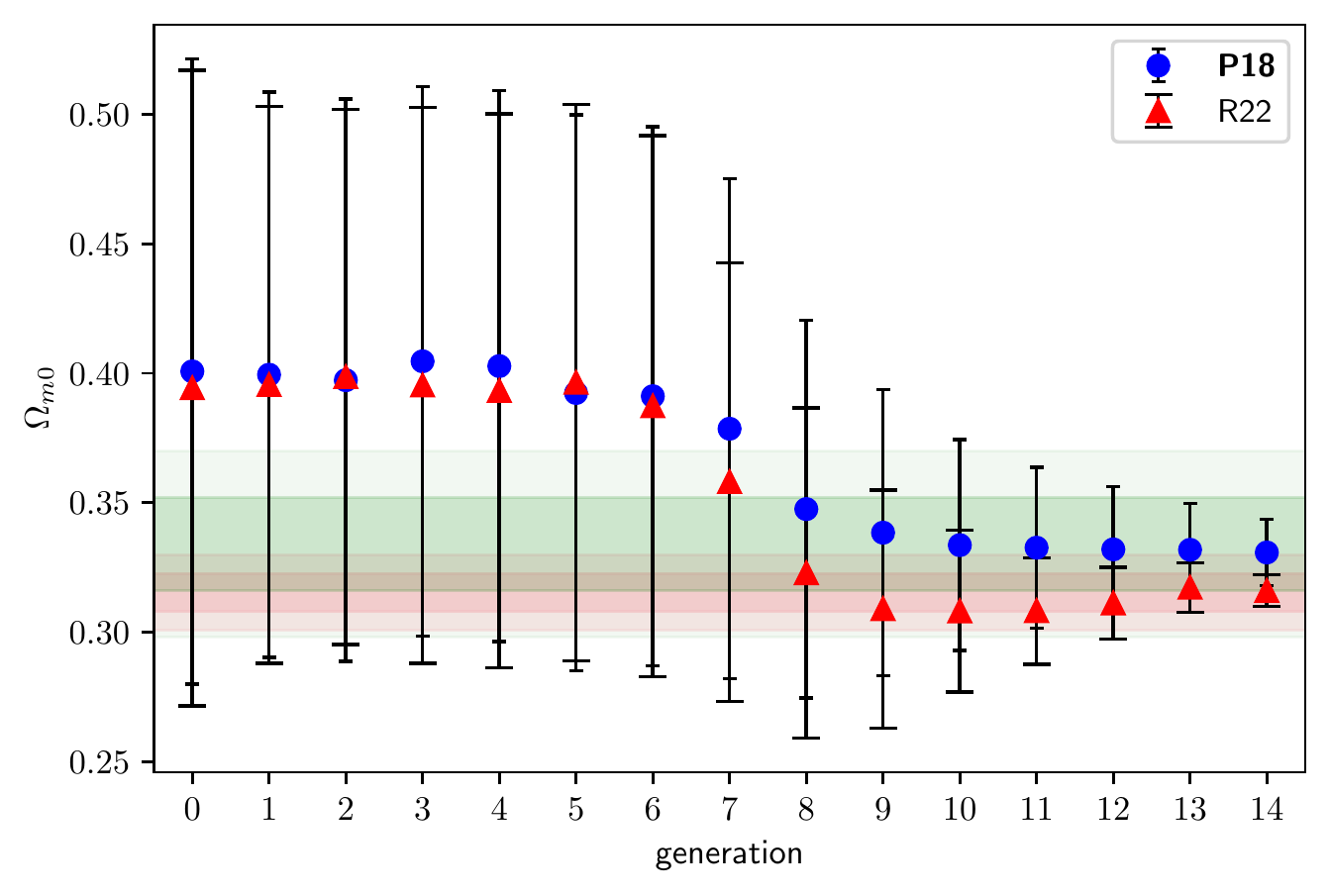}}
    \subfigure[\ \ {SNe + CC + BAO} ]{
		\includegraphics[width = 0.5 \textwidth]{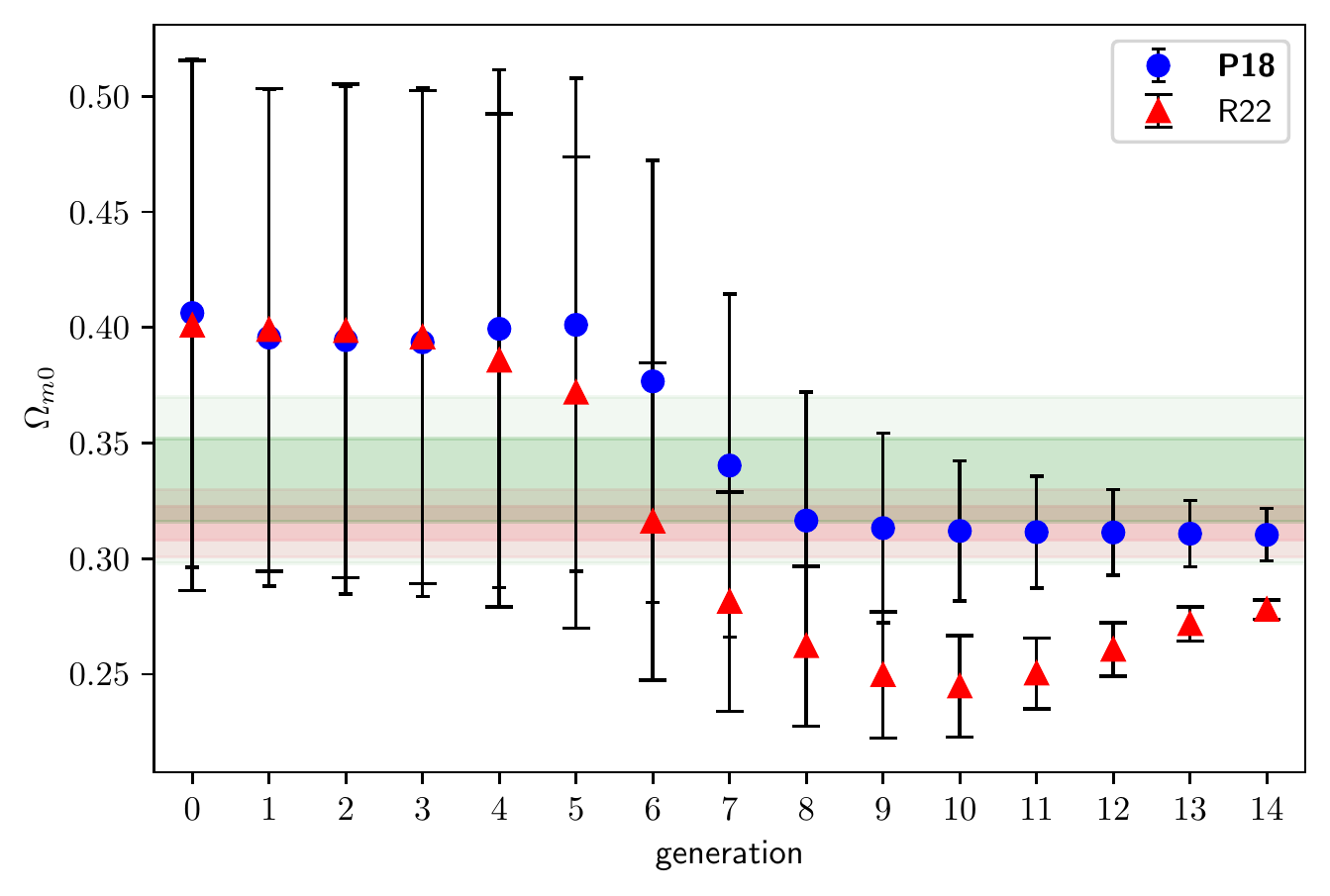}}
    \subfigure[\ \ {SNe + CC + BAO + RSD} ]{
		\includegraphics[width = 0.5 \textwidth]{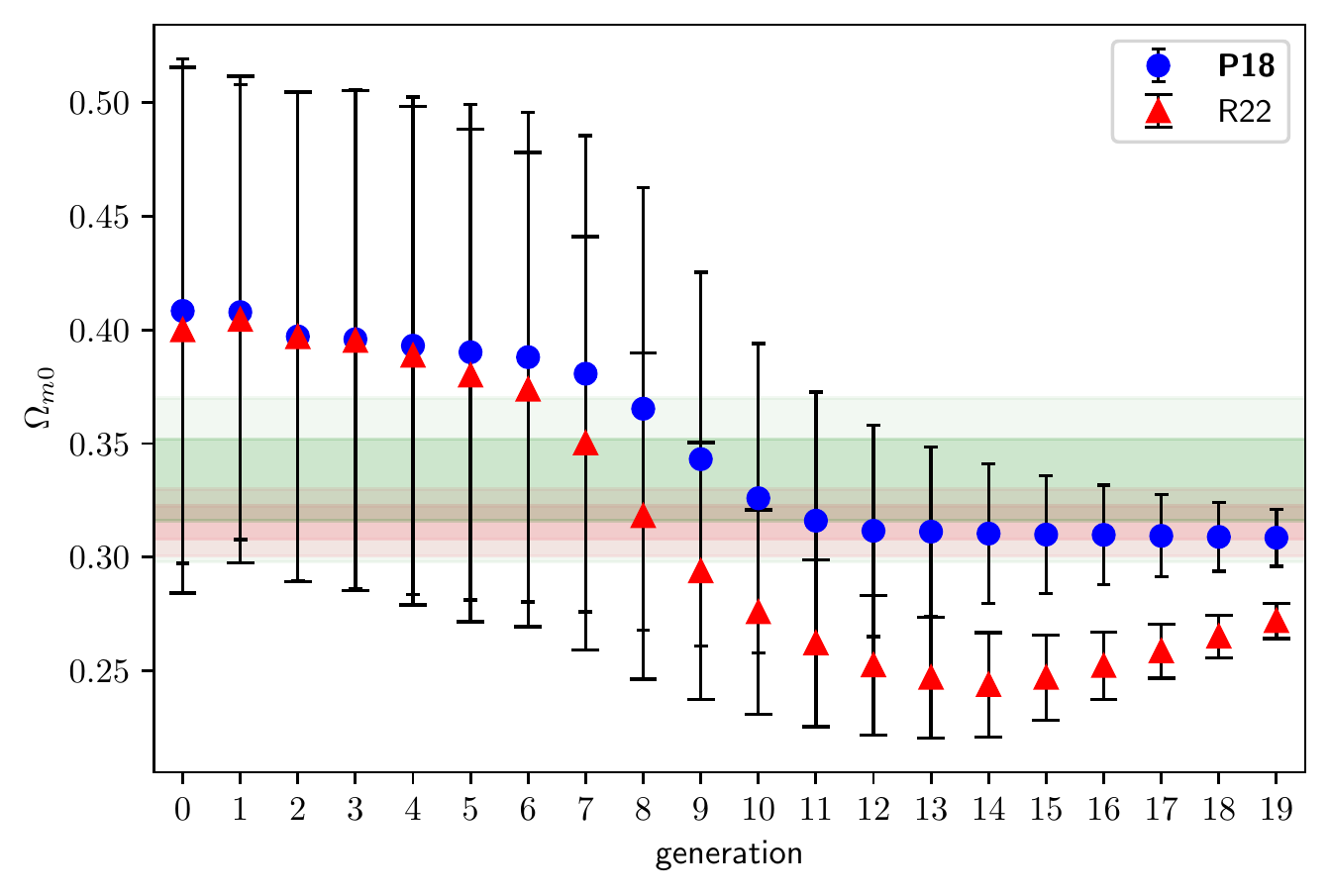}}
    \caption{{
    Matter densities constrained by ABC. The pink and green bands represent constraints from the CMB ($\Omega_{m0} = 0.3153 \pm 0.0073$ {\protect\cite{Aghanim:2018eyx}}) and local observations ($\Omega_{m0} = 0.334 \pm 0.018$ {\protect\cite{Brout:2022vxf}}).
    }}
    \label{fig:matterABC}
\end{figure}

{
In this section, we briefly discuss the matter densities that were constrained simultaneously by ABC. This adds extra significance to the results, particularly that the ABC appears to end up with values consistent with CMB constraints. Figure \ref{fig:matterABC} shows the corresponding matter density evolution plots that were obtained simultaneously with the $H_0$ and $\sigma_{8}$ that were presented in the body of the paper.
}

{
Like the Hubble constant, we see that the matter density starts with conservative priors at early generations. But for this case, there are no tensions to begin with for both models as the error bars are too large. As generations pass on, however, the parameter spaces of each model tighten to keep up with the evolution posed by ABC and we find that eventually a tension grows between the matter densities in P18 and R22. It is interesting to point out that at later generations the R22 model leads to low $\Omega_{m0}$ priors while the P18 gets to higher ones. We understand this is due to the high redshift expansion rate data (CC and BAO) that constrains more directly $\Omega_{m0} h^2$ rather than $\Omega_{m0}$ itself. For this reason, the tension persists more strongly with the BAO, which brings tighter constraints on the expansion. The takeaway we wish to impart regardless is that the naturally selected model (P18) comes with estimates that are consistent with CMB constraints (e.g., pink band in Figure \ref{fig:matterABC}). The matter density evolution brings more support to this statement.
}

\section{{Approximate Bayesian Computation-Sequential Monte Carlo}}
\label{sec:abcsmc}

{
The ABC-SMC algorithm in its full glory \cite{2009arXiv0901.1925T}:
\begin{enumerate}
    \item[1.] Initialize tolerances $\{\epsilon_t\}$ such that $\epsilon_{t + 1} < \epsilon_{t}$ and set population indicator $t = 0$;
    \item[2.] \begin{enumerate}
        \item[0] Set particle indicator $i = 1$;
        \item[1] Sample $m^*$ from $\pi(m)$;
        \begin{itemize}
            \item[] If $t = 0$, sample $\theta^{**}$ from $\pi\left(\theta\left(m^*\right)\right)$.
            \item[] If $t > 0$, sample $\theta^*$ from previous population $\{ \theta\left(m^*\right)_{t-1} \}$ with weights $w\left(m^*\right)_{t-1}$.
            \item[] Perturb the particle $\theta^*$ to obtain $\theta^{**} \sim K_t\left(\theta|\theta^*\right)$.
            \item[] If $\pi\left(\theta^{**}\right) = 0$, return to 2.1.
            \item[] Simulate a candidate data set $x^* \sim f\left(x|\theta^{**},m^{**}\right)$.
            \item[] If $d\left(x^*, x_0\right) \geq \epsilon_t$, return to 2.1.
        \end{itemize}
        \item[2] Set $m_t^{i} = m^*$ and add $\theta^{**}$ to the population of particles $\{ \theta\left(m^*\right)_t \}$, and calculate its weight as
        \begin{equation} \nonumber
            w_t^{(i)} =
            \begin{cases}
            1 & , {\rm if} \ t = 0 \\
            \dfrac{\pi\left( \theta^{**} \right)}{\sum_{j=1}^N w_{t-1}^{(j)} K_t\left( \theta_{t-1}^{j}, \theta^{**} \right) } & , {\rm otherwise} \,.
            \end{cases}
        \end{equation}
        \begin{itemize}
            \item[] If $i < N$ set $i = i + 1$, go to 2.1. 
        \end{itemize}
    \end{enumerate}
    \item[3.] For every $m$, normalize the weights;
    \begin{itemize}
        \item[] For every $t < T$, set $t = t + 1$, go to 2.0.
    \end{itemize}
\end{enumerate}
Above, $\epsilon_i$ are tolerances, $m$ are models, $\theta$ are model parameters, $\pi\left(x\right)$ are prior parameters, $t$ is a generation label, $d(a,b)$ is a distance function, $w$ are weights, and $K_t(\theta_1|\theta_t)$ is a perturbation kernel. {The $\pi(m)$ in this context stands for the initial prior probabilities over the joint model space, e.g., $\pi(m_1) = 0.3, \pi(m_2) = 0.2, \cdots$ for models $m_i$. This is reasonably often taken to just be flat, as in our case, such that $\pi({\rm P18}) = \pi({\rm R22}) = 0.5$ during the calibration sample or generation $t = -1$.} See \cite{2009arXiv0901.1925T, 2009arXiv0911.1705T, 2009arXiv0910.4472T} for more details and examples.
}



\end{document}